**NISTIR 8202**

# Blockchain Technology Overview

Dylan Yaga
Peter Mell
Nik Roby
Karen Scarfone

**NIST**

**National Institute of
Standards and Technology**

U.S. Department of Commerce

# Blockchain Technology Overview

Dylan Yaga
Peter Mell
*Computer Security Division*
*Information Technology Laboratory*

Nik Roby
*G2, Inc.*
*Annapolis Junction, MD*

Karen Scarfone
*Scarfone Cybersecurity*
*Clifton, VA*

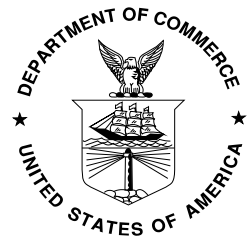

**Comments on this publication may be submitted to:**

All comments are subject to release under the Freedom of Information Act (FOIA).

## Reports on Computer Systems Technology

The Information Technology Laboratory (ITL) at the National Institute of Standards and Technology (NIST) promotes the U.S. economy and public welfare by providing technical leadership for the Nation's measurement and standards infrastructure. ITL develops tests, test methods, reference data, proof of concept implementations, and technical analyses to advance the development and productive use of information technology. ITL's responsibilities include the development of management, administrative, technical, and physical standards and guidelines for the cost-effective security and privacy of other than national security-related information in federal information systems.

## Abstract

Blockchains are tamper evident and tamper resistant digital ledgers implemented in a distributed fashion (i.e., without a central repository) and usually without a central authority (i.e., a bank, company, or government). At their basic level, they enable a community of users to record transactions in a shared ledger within that community, such that under normal operation of the blockchain network no transaction can be changed once published. This document provides a high-level technical overview of blockchain technology. The purpose is to help readers understand how blockchain technology works.

## Keywords

## Acknowledgments

The authors wish to thank all contributors to this publication, and their colleagues who reviewed drafts of this report and contributed technical and editorial additions. This includes NIST staff James Dray, Sandy Ressler, Rick Kuhn, Lee Badger, Eric Trapnell, Mark Trapnell, James Shook and Michael Davidson.

Additional thanks to all the people and organizations who submitted comments during the public comment period.

## Audience

This publication is designed for readers with little or no knowledge of blockchain technology who wish to understand at a high level how it works. It is not intended to be a technical guide; the discussion of the technology provides a conceptual understanding. Note that some examples, figures, and tables are simplified to fit the audience.

## Trademark Information

All registered trademarks and trademarks belong to their respective organizations.

## Executive Summary

Blockchains are tamper evident and tamper resistant digital ledgers implemented in a distributed fashion (i.e., without a central repository) and usually without a central authority (i.e., a bank, company, or government). At their basic level, they enable a community of users to record transactions in a shared ledger within that community, such that under normal operation of the blockchain network no transaction can be changed once published. In 2008, the blockchain idea was combined with several other technologies and computing concepts to create modern cryptocurrencies: electronic cash protected through cryptographic mechanisms instead of a central repository or authority. The first such blockchain based cryptocurrency was Bitcoin.

Within the Bitcoin blockchain, information representing electronic cash is attached to a digital address. Bitcoin users can digitally sign and transfer rights to that information to another user and the Bitcoin blockchain records this transfer publicly, allowing all participants of the network to independently verify the validity of the transactions. The Bitcoin blockchain is stored, maintained, and collaboratively managed by a distributed group of participants. This, along with certain cryptographic mechanisms, makes the blockchain resilient to attempts to alter the ledger later (modifying blocks or forging transactions).

Because there are countless news articles and videos describing the "magic" of blockchain technology, this paper aims to describe the method behind the magic (i.e., how blockchain technology works). Arthur C. Clarke once wrote, "Any sufficiently advanced technology is indistinguishable from magic" [1]. Clarke's statement is a perfect representation for the emerging applications of blockchain technology. There is hype around the use of blockchain technology, yet the technology is not well understood. It is not magical; it will not solve all problems. As with all new technology, there is a tendency to want to apply it to every sector in every way imaginable. To help promote correct application, this document provides information necessary to develop a high-level understanding of the technology.

Blockchain technology is the foundation of modern cryptocurrencies, so named because of the heavy usage of cryptographic functions. Users utilize public and private keys to digitally sign and securely transact within the system. For cryptocurrency based blockchain networks which utilize mining (see section 4.1), users may solve puzzles using cryptographic hash functions in hopes of being rewarded with a fixed amount of the cryptocurrency. However, blockchain technology may be more broadly applicable than cryptocurrencies. In this work, we focus on the cryptocurrency use case, since that is the primary use of the technology today; however, there is a growing interest in other sectors.

Organizations considering implementing blockchain technology need to understand fundamental aspects of the technology. For example, what happens when an organization implements a blockchain network and then decides they need to make modifications to the data stored? When using a database, modifying the actual data can be accomplished through a database query and update. Organizations must understand that while changes to the actual blockchain data may be difficult, applications using the blockchain as a data layer work around this by treating later blocks and transactions as updates or modifications to earlier blocks and transactions. This software abstraction allows for modifications to working data, while providing a full history of changes. Another critical aspect of blockchain technology is how the participants agree that a transaction is valid. This is called "reaching consensus", and there are many models for doing so, each with positives and negatives for particular business cases. It is important to understand that a blockchain is just one part of a solution.

Blockchain implementations are often designed with a specific purpose or function. Example functions include cryptocurrencies, smart contracts (software deployed on the blockchain and executed by computers running that blockchain), and distributed ledger systems between businesses. There has been a constant stream of developments in the field of blockchain technology, with new platforms being announced constantly – the landscape is continuously changing.

There are two general high-level categories for blockchain approaches that have been identified: permissionless, and permissioned. In a permissionless blockchain network anyone can read and write to the blockchain without authorization. Permissioned blockchain networks limit participation to specific people or organizations and allow finer-grained controls. Knowing the differences between these two categories allows an organization to understand which subset of blockchain technologies may be applicable to its needs.

Despite the many variations of blockchain networks and the rapid development of new blockchain related technologies, most blockchain networks use common core concepts. Blockchains are a distributed ledger comprised of blocks. Each block is comprised of a block header containing metadata about the block, and block data containing a set of transactions and other related data. Every block header (except for the very first block of the blockchain) contains a cryptographic link to the previous block's header. Each transaction involves one or more blockchain network users and a recording of what happened, and it is digitally signed by the user who submitted the transaction.

Blockchain technology takes existing, proven concepts and merges them together into a single solution. This document explores the fundamentals of how these technologies work and the differences between blockchain approaches. This includes how the participants in the network come to agree on whether a transaction is valid and what happens when changes need to be made to an existing blockchain deployment. Additionally, this document explores when to consider using a blockchain network.

The use of blockchain technology is not a silver bullet, and there are issues that must be considered such as how to deal with malicious users, how controls are applied, and the limitations of the implementations. Beyond the technology issues that need to be considered, there are operational and governance issues that affect the behavior of the network. For example, in permissioned blockchain networks, described later in this document, there are design issues surrounding what entity or entities will operate and govern the network for the intended user base.

Blockchain technology is still new and should be investigated with the mindset of "how could blockchain technology potentially benefit us?" rather than "how can we make our problem fit into the blockchain technology paradigm?". Organizations should treat blockchain technology like they would any other technological solution at their disposal and use it in appropriate situations.

# Table of Contents

## List of Appendices

## List of Tables and Figures

## 1       Introduction

Blockchains are tamper evident and tamper resistant digital ledgers implemented in a distributed fashion (i.e., without a central repository) and usually without a central authority (i.e., a bank, company or government). At their basic level, they enable a community of users to record transactions in a shared ledger within that community, such that under normal operation of the blockchain network no transaction can be changed once published. In 2008, the blockchain idea was combined with several other technologies and computing concepts to create modern cryptocurrencies: electronic cash protected through cryptographic mechanisms instead of a central repository or authority.

This technology became widely known in 2009 with the launch of the Bitcoin network, the first of many modern cryptocurrencies. In Bitcoin, and similar systems, the transfer of digital information that represents electronic cash takes place in a distributed system. Bitcoin users can digitally sign and transfer their rights to that information to another user and the Bitcoin blockchain records this transfer publicly, allowing all participants of the network to independently verify the validity of the transactions. The Bitcoin blockchain is independently maintained and managed by a distributed group of participants. This, along with cryptographic mechanisms, makes the blockchain resilient to attempts to alter the ledger later (modifying blocks or forging transactions). Blockchain technology has enabled the development of many cryptocurrency systems such as Bitcoin and Ethereum[1]. Because of this, blockchain technology is often viewed as bound to Bitcoin or possibly cryptocurrency solutions in general. However, the technology is available for a broader variety of applications and is being investigated for a variety of sectors.

The numerous components of blockchain technology along with its reliance on cryptographic primitives and distributed systems can make it challenging to understand. However, each component can be described simply and used as a building block to understand the larger complex system. Blockchains can be informally defined as:

> Blockchains are distributed digital ledgers of cryptographically signed transactions that are grouped into blocks. Each block is cryptographically linked to the previous one (making it tamper evident) after validation and undergoing a consensus decision. As new blocks are added, older blocks become more difficult to modify (creating tamper resistance). New blocks are replicated across copies of the ledger within the network, and any conflicts are resolved automatically using established rules.

---

[1] Bitcoin and Ethereum are mentioned here since they are listed as the top two cryptocurrencies on market capitalization websites

## 1.1 Background and History

The core ideas behind blockchain technology emerged in the late 1980s and early 1990s. In 1989, Leslie Lamport developed the Paxos protocol, and in 1990 submitted the paper *The Part-Time Parliament* [2] to ACM Transactions on Computer Systems; the paper was finally published in a 1998 issue. The paper describes a consensus model for reaching agreement on a result in a network of computers where the computers or network itself may be unreliable. In 1991, a signed chain of information was used as an electronic ledger for digitally signing documents in a way that could easily show none of the signed documents in the collection had been changed [3]. These concepts were combined and applied to electronic cash in 2008 and described in the paper, *Bitcoin: A Peer to Peer Electronic Cash System* [4], which was published pseudonymously by Satoshi Nakamoto, and then later in 2009 with the establishment of the Bitcoin cryptocurrency blockchain network. Nakamoto's paper contained the blueprint that most modern cryptocurrency schemes follow (although with variations and modifications). Bitcoin was just the first of many blockchain applications.

Many electronic cash schemes existed prior to Bitcoin (e.g., ecash and NetCash), but none of them achieved widespread use. The use of a blockchain enabled Bitcoin to be implemented in a distributed fashion such that no single user controlled the electronic cash and no single point of failure existed; this promoted its use. Its primary benefit was to enable direct transactions between users without the need for a trusted third party. It also enabled the issuance of new cryptocurrency in a defined manner to those users who manage to publish new blocks and maintain copies of the ledger; such users are called *miners* in Bitcoin. The automated payment of the miners enabled distributed administration of the system without the need to organize. By using a blockchain and consensus-based maintenance, a self-policing mechanism was created that ensured that only valid transactions and blocks were added to the blockchain.

In Bitcoin, the blockchain enabled users to be pseudonymous. This means that users are anonymous, but their account identifiers are not; additionally, all transactions are publicly visible. This has effectively enabled Bitcoin to offer pseudo-anonymity because accounts can be created without any identification or authorization process (such processes are typically required by Know-Your-Customer (KYC) laws).

Since Bitcoin was pseudonymous, it was essential to have mechanisms to create trust in an environment where users could not be easily identified. Prior to the use of blockchain technology, this trust was typically delivered through intermediaries trusted by both parties. Without trusted intermediaries, the needed trust within a blockchain network is enabled by four key characteristics of blockchain technology, described below:

- **Ledger** – the technology uses an append only ledger to provide full transactional history. Unlike traditional databases, transactions and values in a blockchain are not overridden.
- **Secure** – blockchains are cryptographically secure, ensuring that the data contained within the ledger has not been tampered with, and that the data within the ledger is attestable.
- **Shared** – the ledger is shared amongst multiple participants. This provides transparency across the node participants in the blockchain network.

- **Distributed** – the blockchain can be distributed. This allows for scaling the number of nodes of a blockchain network to make it more resilient to attacks by bad actors. By increasing the number of nodes, the ability for a bad actor to impact the consensus protocol used by the blockchain is reduced.

For blockchain networks that allow anyone to anonymously create accounts and participate (called permissionless blockchain networks), these capabilities deliver a level of trust amongst parties with no prior knowledge of one another; this trust can enable individuals and organizations to transact directly, which may result in transactions being delivered faster and at lower costs. For a blockchain network that more tightly controls access (called permissioned blockchain networks), where some trust may be present among users, these capabilities help to bolster that trust.

## 1.2   Purpose and Scope

This document provides a high-level technical overview of blockchain technology. It looks at different categories of implementation approaches. It discusses the components of blockchain technology and provides diagrams and examples when possible. It discusses, at a high-level, some consensus models used in blockchain networks. It also provides an overview of how blockchain technology changes (known as forking) affect the blockchain network. It provides details on how blockchain technology was extended beyond attestable transactions to include attestable application processes known as smart contracts. It also touches on some of the limitations and misconceptions surrounding the technology. Finally, this document presents several areas that organizations should consider when investigating blockchain technology. It is intended to help readers to understand the technologies which comprise blockchain networks.

## 1.3   Notes on Terms

The terminology for blockchain technology varies from one implementation to the next – to talk about the technology, generic terms will be used. Throughout this document the following terms will be used:

- *Blockchain* – the actual ledger
- *Blockchain technology* – a term to describe the technology in the most generic form
- *Blockchain network* – the network in which a blockchain is being used
- *Blockchain implementation* – a specific blockchain
- *Blockchain network user* – a person, organization, entity, business, government, etc. which is utilizing the blockchain network
- *Node* – an individual system within a blockchain network
  - *Full node* – a node that stores the entire blockchain, ensures transactions are valid
    - *Publishing node* – a full node that also publishes new blocks
  - *Lightweight node* – a node that does not store or maintain a copy of the blockchain and must pass their transactions to full nodes

## 1.4    Results of the Public Comment Period

This document has seen substantial revision in response to the public comments received. Part of the revising process was to tighten the scope, and to provide a more foundational document as an introduction to the technology. Please note that several sections present in the draft (7.1.2 - Permissioned Use Cases, 7.2.2 - Permissionless Use Cases, and 8 - Blockchain Platforms) are not present in the published version. These topics were made explicitly out of scope for this document because the rapidly changing landscape and areas of interest around this technology, as well as the ever-increasing number of platforms, would make these sections out of place in such a foundational document. The topics in these sections are still being considered for future works.

Additionally, section 8.1.2 – Bitcoin Cash contained an erroneous and unverified statement which was not identified and removed during initial editing of the draft. Since this section has been removed, this issue is now addressed.

## 1.5    Document Structure

The rest of this document is organized as follows:

- **Section 2** discusses the high-level categorization of blockchain technology: permissionless and permissioned.
- **Section 3** defines the high-level components of a blockchain network architecture, including hashes, transactions, ledgers, blocks, and blockchains.
- **Section 4** discusses several consensus models employed by blockchain technology.
- **Section 5** introduces the concept of forking.
- **Section 6** discusses smart contracts.
- **Section 7** discusses several limitations as well as misconceptions surrounding blockchain technology.
- **Section 8** discusses various application considerations, as well as provides additional considerations from government, academia, and technology enthusiasts.
- **Section 9** is the conclusion.
- **Appendix A** provides a list of acronyms and abbreviations used in the document.
- **Appendix B** contains a glossary for selected terms defined in the document.
- **Appendix C** lists the references used throughout the document.

## 2    Blockchain Categorization

Blockchain networks can be categorized based on their permission model, which determines who can maintain them (e.g., publish blocks). If anyone can publish a new block, it is *permissionless*. If only particular users can publish blocks, it is *permissioned*. In simple terms, a permissioned blockchain network is like a corporate intranet that is controlled, while a permissionless blockchain network is like the public internet, where anyone can participate. Permissioned blockchain networks are often deployed for a group of organizations and individuals, typically referred to as a consortium. This distinction is necessary to understand as it impacts some of the blockchain components discussed later in this document.

### 2.1    Permissionless

Permissionless blockchain networks are decentralized ledger platforms open to anyone publishing blocks, without needing permission from any authority. Permissionless blockchain platforms are often open source software, freely available to anyone who wishes to download them. Since anyone has the right to publish blocks, this results in the property that anyone can read the blockchain as well as issue transactions on the blockchain (through including those transactions within published blocks). Any blockchain network user within a permissionless blockchain network can read and write to the ledger. Since permissionless blockchain networks are open to all to participate, malicious users may attempt to publish blocks in a way that subverts the system (discussed in detail later). To prevent this, permissionless blockchain networks often utilize a multiparty agreement or 'consensus' system (see Section 4) that requires users to expend or maintain resources when attempting to publish blocks. This prevents malicious users from easily subverting the system. Examples of such consensus models include proof of work (see Section 4.1) and proof of stake (see Section 4.2) methods. The consensus systems in permissionless blockchain networks usually promote non-malicious behavior through rewarding the publishers of protocol-conforming blocks with a native cryptocurrency.

### 2.2    Permissioned

Permissioned blockchain networks are ones where users publishing blocks must be authorized by some authority (be it centralized or decentralized). Since only authorized users are maintaining the blockchain, it is possible to restrict read access and to restrict who can issue transactions. Permissioned blockchain networks may thus allow anyone to read the blockchain or they may restrict read access to authorized individuals. They also may allow anyone to submit transactions to be included in the blockchain or, again, they may restrict this access only to authorized individuals. Permissioned blockchain networks may be instantiated and maintained using open source or closed source software.

Permissioned blockchain networks can have the same traceability of digital assets as they pass through the blockchain, as well as the same distributed, resilient, and redundant data storage system as a permissionless blockchain networks. They also use consensus models for publishing blocks, but these methods often do not require the expense or maintenance of resources (as is the case with current permissionless blockchain networks). This is because the establishment of one's identity is required to participate as a member of the permissioned blockchain network; those maintaining the blockchain have a level of trust with each other, since they were all authorized to publish blocks and since their authorization can be revoked if they misbehave. Consensus models in permissioned blockchain networks are then usually faster and less computationally expensive.

Permissioned blockchain networks may also be used by organizations that need to more tightly control and protect their blockchain. However, if a single entity controls who can publish blocks, the users of the blockchain will need to have trust in that entity. Permissioned blockchain networks may also be used by organizations that wish to work together but may not fully trust one another. They can establish a permissioned blockchain network and invite business partners to record their transactions on a shared distributed ledger. These organizations can determine the consensus model to be used, based on how much they trust one another. Beyond trust, permissioned blockchain networks provide transparency and insight that may help better inform business decisions and hold misbehaving parties accountable. This can explicitly include auditing and oversight entities making audits a constant occurrence versus a periodic event.

Some permissioned blockchain networks support the ability to selectively reveal transaction information based on a blockchain network users identity or credentials. With this feature, some degree of privacy in transactions may be obtained. For example, it could be that the blockchain records that a transaction between two blockchain network users took place, but the actual contents of transactions is only accessible to the involved parties.

Some permissioned blockchain networks require all users to be authorized to send and receive transactions (they are not anonymous, or even pseudo-anonymous). In such systems parties work together to achieve a shared business process with natural disincentives to commit fraud or otherwise behave as a bad actor (since they can be identified). If bad behavior were to occur, it is well known where the organizations are incorporated, what legal remedies are available and how to pursue those remedies in the relevant judicial system.

## 3      Blockchain Components

Blockchain technology can seem complex; however, it can be simplified by examining each component individually. At a high level, blockchain technology utilizes well-known computer science mechanisms and cryptographic primitives (cryptographic hash functions, digital signatures, asymmetric-key cryptography) mixed with record keeping concepts (such as append only ledgers). This section discusses each individual main component: cryptographic hash functions, transactions, asymmetric-key cryptography, addresses, ledgers, blocks, and how blocks are chained together.

### 3.1    Cryptographic Hash Functions

An important component of blockchain technology is the use of cryptographic hash functions for many operations. *Hashing* is a method of applying a cryptographic hash function to data, which calculates a relatively unique output (called a *message digest*, or just *digest*) for an input of nearly any size (e.g., a file, text, or image). It allows individuals to independently take input data, hash that data, and derive the same result – proving that there was no change in the data. Even the smallest change to the input (e.g., changing a single bit) will result in a completely different output digest. Table 1 shows simple examples of this.

Cryptographic hash functions have these important security properties:

1. They are *preimage resistant.* This means that they are one-way; it is computationally infeasible to compute the correct input value given some output value (e.g., given a digest, find $x$ such that hash($x$) = digest).
2. They are *second preimage resistant.* This means one cannot find an input that hashes to a specific output. More specifically, cryptographic hash functions are designed so that given a specific input, it is computationally infeasible to find a second input which produces the same output (e.g., given $x$, find $y$ such that hash($x$) = hash($y$)). The only approach available is to exhaustively search the input space, but this is computationally infeasible to do with any chance of success.
3. They are *collision resistant.* This means that one cannot find two inputs that hash to the same output. More specifically, it is computationally infeasible to find any two inputs that produce the same digest (e.g., find an $x$ and $y$ which hash($x$) = hash($y$)).

A specific cryptographic hash function used in many blockchain implementations is the Secure Hash Algorithm (SHA) with an output size of 256 bits (SHA-256). Many computers support this algorithm in hardware, making it fast to compute. SHA-256 has an output of 32 bytes (1 byte = 8 bits, 32 bytes = 256 bits), generally displayed as a 64-character hexadecimal string (see Table 1 below).

This means that there are $2^{256} \approx 10^{77}$, or 115,792,089,237,316,195,423,570,985,008,687,907,853,269,984,665,640,564,039,457,584,007,913,129,639,936 possible digest values. The algorithm for SHA-256, as well as others, is specified in Federal Information Processing Standard (FIPS) 180-4 [5]. The NIST Secure Hashing website [6] contains FIPS specifications for all NIST-approved hashing algorithms.

**Table 1: Examples of Input Text and Corresponding SHA-256 Digest Values**

| Input Text | SHA-256 Digest Value |
|---|---|
| 1 | 0x6b86b273ff34fce19d6b804eff5a3f5747ada4eaa22f1d49c01e52ddb7875b4b |
| 2 | 0xd4735e3a265e16eee03f59718b9b5d03019c07d8b6c51f90da3a666eec13ab35 |
| Hello, World! | 0xdffd6021bb2bd5b0af676290809ec3a53191dd81c7f70a4b28688a362182986f |

Since there are an infinite number of possible input values and a finite number of possible output digest values, it is possible but highly unlikely to have a collision where hash($x$) = hash($y$) (i.e., the hash of two different inputs produces the same digest). SHA-256 is said to be collision resistant, since to find a collision in SHA-256, one would have to execute the algorithm, on average, about $2^{128}$ times (which is 340 undecillions, or more precisely 340,282,366,920,938,463,463,374,607,431,768,211,456; roughly 3.402 x $10^{38}$).

To put this into perspective, the hash rate (hashes per second) of the entire Bitcoin network in 2015 was 300 quadrillion hashes per second (300,000,000,000,000,000/s) [7]. At that rate, it would take the entire Bitcoin network roughly 35,942,991,748,521 (roughly 3.6 x $10^{13}$) years[2] to manufacture a collision (note that the universe is estimated to be 1.37 x $10^{10}$ years old)[3]. Even if any such input $x$ and $y$ that produce the same digest, it would be also very unlikely for both inputs to be valid in the context of the blockchain network (i.e., $x$ and $y$ are both valid transactions).

Within a blockchain network, cryptographic hash functions are used for many tasks, such as:

- Address derivation – discussed in section 3.4.
- Creating unique identifiers.
- Securing the block data – a publishing node will hash the block data, creating a digest that will be stored within the block header.
- Securing the block header – a publishing node will hash the block header. If the blockchain network utilizes a proof of work consensus model (see Section 4.1), the publishing node will need to hash the block header with different nonce values (see Section 3.1.1) until the puzzle requirements have been fulfilled. The current block header's hash digest will be included within the next block's header, where it will secure the current block header data.

Because the block header includes a hash representation of the block data, the block data itself is also secured when the block header digest is stored in the next block.

---

[2] Calculation: $2^{128}$/((((300000000000000000×60) (hash per second -> minute)
    ×60) (minute -> hour)
    ×24) (hour -> day)
    ×365.25) (day -> year) = 35942991748521.06026898693261758057345467758426918193 years
    https://www.wolframalpha.com/input/?i=2%5E128%2F(30000000000000000+*+60+*+60+*+24+*+365.25)

[3] As estimated by measurements made by the Wilkinson Microwave Anisotropy Probe
    https://map.gsfc.nasa.gov/universe/uni_age.html

There are many families of cryptographic hash functions utilized in blockchain technology (SHA-256 is not the only one), such as Keccak (which was selected by NIST as the winner of a competition to create the SHA-3 hashing standard), as well as RIPEMD-160.[8]

### 3.1.1 Cryptographic Nonce

A cryptographic nonce is an arbitrary number that is only used once. A cryptographic nonce can be combined with data to produce different hash digests per nonce:

$$hash \ (data + nonce) = digest$$

Only changing the nonce value provides a mechanism for obtaining different digest values while keeping the same data. This technique is utilized in the proof of work consensus model (see Section 4.1).

### 3.2 Transactions

A *transaction* represents an interaction between parties. With cryptocurrencies, for example, a transaction represents a transfer of the cryptocurrency between blockchain network users. For business-to-business scenarios, a transaction could be a way of recording activities occurring on digital or physical assets.  Figure 1 shows a notional example of a cryptocurrency transaction. Each block in a blockchain can contain zero or more transactions. For some blockchain implementations, a constant supply of new blocks (even with zero transactions) is critical to maintain the security of the blockchain network; by having a constant supply of new blocks being published, it prevents malicious users from ever "catching up" and manufacturing a longer, altered blockchain (see Section 4.7).

The data which comprises a transaction can be different for every blockchain implementation, however the mechanism for transacting is largely the same. A blockchain network user sends information to the blockchain network. The information sent may include the sender's address (or another relevant identifier), sender's public key, a digital signature, transaction inputs and transaction outputs.

A single cryptocurrency transaction typically requires at least the following information, but can contain more:

- **Inputs** – The inputs are usually a list of the digital assets to be transferred. A transaction will reference the source of the digital asset (providing provenance) – either the previous transaction where it was given to the sender, or for the case of new digital assets, the origin event. Since the input to the transaction is a reference to past events, the digital assets do not change. In the case of cryptocurrencies this means that value cannot be added or removed from existing digital assets. Instead, a single digital asset can be split into multiple new digital assets (each with lesser value) or multiple digital assets can be combined to form fewer new digital assets (with a correspondingly greater value). The splitting or joining of assets will be specified within the transaction output.

The sender must also provide proof that they have access to the referenced inputs, generally by digitally signing the transaction – proving access to the private key.

- **Outputs** – The outputs are usually the accounts that will be the recipients of the digital assets along with how much digital asset they will receive. Each output specifies the number of digital assets to be transferred to the new owner(s), the identifier of the new owner(s), and a set of conditions the new owners must meet to spend that value. If the digital assets provided are more than required, the extra funds must be explicitly sent back to the sender (this is a mechanism to "make change").

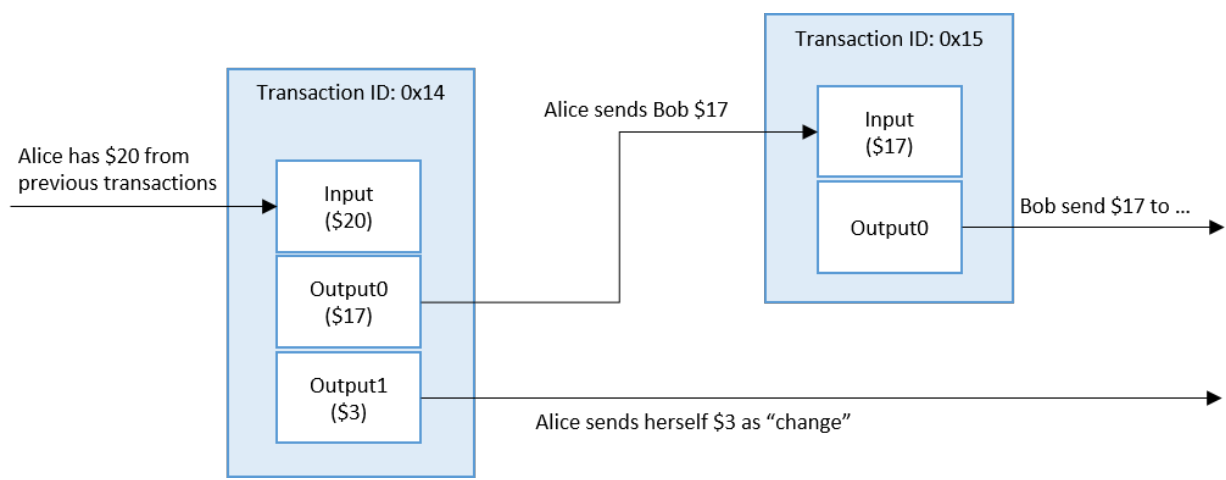

**Figure 1 - Example Cryptocurrency Transaction**

While primarily used to transfer digital assets, transactions can be more generally used to transfer data. In a simple case, someone may simply want to permanently and publicly post data on the blockchain. In the case of smart contract systems, transactions can be used to send data, process that data, and store some result on the blockchain. For example, a transaction can be used to change an attribute of a digitized asset such as the location of a shipment within a blockchain technology-based supply chain system.

Regardless of how the data is formed and transacted, determining the validity and authenticity of a transaction is important. The validity of a transaction ensures that the transaction meets the protocol requirements and any formalized data formats or smart contract requirements specific to the blockchain implementation. The authenticity of a transaction is also important, as it determines that the sender of digital assets had access to those digital assets. Transactions are typically digitally signed by the sender's associated private key (asymmetric-key cryptography is briefly discussed in Section 3.3) and can be verified at any time using the associated public key.

## 3.3   Asymmetric-Key Cryptography

Blockchain technology uses asymmetric-key cryptography[4] (also referred to as public key cryptography). Asymmetric-key cryptography uses a pair of keys: a public key and a private key that are mathematically related to each other. The public key is made public without reducing the security of the process, but the private key must remain secret if the data is to retain its cryptographic protection. Even though there is a relationship between the two keys, the private key cannot efficiently be determined based on knowledge of the public key. One can encrypt with a private key and then decrypt with the public key. Alternately, one can encrypt with a public key and then decrypt with a private key.

Asymmetric-key cryptography enables a trust relationship between users who do not know or trust one another, by providing a mechanism to verify the integrity and authenticity of transactions while at the same time allowing transactions to remain public.  To do this, the transactions are 'digitally signed'. This means that a private key is used to encrypt a transaction such that anyone with the public key can decrypt it. Since the public key is freely available, encrypting the transaction with the private key proves that the signer of the transaction has access to the private key. Alternately, one can encrypt data with a user's public key such that only users with access to the private key can decrypt it. A drawback is that asymmetric-key cryptography is often slow to compute.

This contrasts with symmetric-key cryptography in which a single secret key is used to both encrypt and decrypt. With symmetric-key cryptography users must already have a trust relationship established with one another to exchange the pre-shared key. In a symmetric system, any encrypted data that can be decrypted with the pre-shared key confirms it was sent by another user with access to the pre-shared key; no user without access to the pre-shared key will be able to view the decrypted data. Compared to asymmetric-key cryptography, symmetric-key cryptography is very fast to compute. Because of this, when one claims to be encrypting something using asymmetric-key cryptography, oftentimes the data is encrypted with symmetric-key cryptography and then the symmetric-key is encrypted using asymmetric-key cryptography. This 'trick' can greatly speed up asymmetric-key cryptography.

Here is a summary of the use of asymmetric-key cryptography in many blockchain networks:

- Private keys are used to digitally sign transactions.
- Public keys are used to derive addresses.
- Public keys are used to verify signatures generated with private keys.
- Asymmetric-key cryptography provides the ability to verify that the user transferring value to another user is in possession of the private key capable of signing the transaction.

---

[4] FIPS Publication 186-4, Digital Signature Standard [9] specifies a common algorithm for digital signing used in blockchain technologies: Elliptic Curve Digital Signature Algorithm (ECDSA).

Some permissioned blockchain networks can leverage a business's existing public key infrastructure for asymmetric-key cryptography to provide user credentials – rather than having each blockchain network user manage their own asymmetric-keys. This is done by utilizing existing directory services and using that information within the blockchain network. Blockchain networks which utilize an existing directory service can access it via existing protocols, such as the Lightweight Directory Access Protocol (LDAP) [10], and utilize the information from the directory natively, or import it into an internal certificate authority within the blockchain network.

## 3.4    Addresses and Address Derivation

Some blockchain networks make use of an *address*, which is a short, alphanumeric string of characters derived from the blockchain network user's public key using a cryptographic hash function, along with some additional data (e.g., version number, checksums). Most blockchain implementations make use of addresses as the "to" and "from" endpoints in a transaction. Addresses are shorter than the public keys and are not secret. One method to generate an address is to create a public key, applying a cryptographic hash function to it, and converting the hash to text:

$$\text{public key} \rightarrow \text{cryptographic hash function} \rightarrow \text{address}$$

Each blockchain implementation may implement a different method to derive an address. For permissionless blockchain networks, which allow anonymous account creation, a blockchain network user can generate as many asymmetric-key pairs, and therefore addresses as desired, allowing for a varying degree of pseudo-anonymity. Addresses may act as the public-facing identifier in a blockchain network for a user, and oftentimes an address will be converted into a QR code (Quick Response Code, a 2-dimensional bar code which can contain arbitrary data) for easier use with mobile devices.

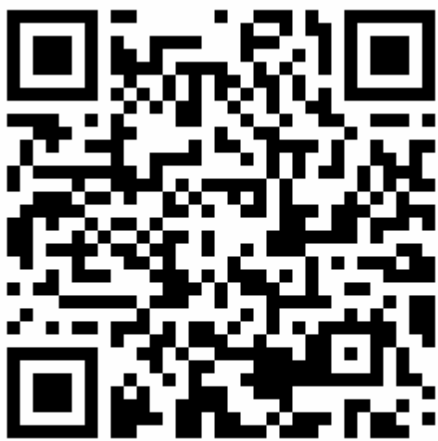

**Figure 2 - A QR code example which has encoded the text "NISTIR 8202 - Blockchain Technology Overview QR code example"**

Blockchain network users may not be the only source of addresses within blockchain networks. It is necessary to provide a method of accessing a smart contract once it has been deployed within a blockchain network. For Ethereum, smart contracts are accessible via a special address called a contract account. This account address is created when a smart contract is deployed (the address for a contract account is deterministically computed from the smart contract creator's address [11]). This contract account allows for the contract to be executed whenever it receives a transaction, as well as create additional smart contracts in turn.

### 3.4.1 Private Key Storage

With some blockchain networks (especially with permissionless blockchain networks), users must manage and securely store their own private keys. Instead of recording them manually, they often use software to securely store them. This software is often referred to as a *wallet*. The wallet can store private keys, public keys, and associated addresses. It may also perform other functions, such as calculating the total number of digital assets a user may have.

If a user loses a private key, then any digital asset associated with that key is lost, because it is computationally infeasible to regenerate the same private key. If a private key is stolen, the attacker will have full access to all digital assets controlled by that private key. The security of private keys is so important that many users use special secure hardware to store them; alternatively, users may take advantage of an emerging industry of private key escrow services. These key escrow services can also satisfy KYC laws in addition to storing private keys as users must provide proof of their identity when creating an account.

Private key storage is an extremely important aspect of blockchain technology. When it is reported in the news that "Cryptocurrency XYZ was stolen from…", it almost certainly means some private keys were found and used to sign a transaction sending the money to a new account, not that the blockchain network itself was compromised. Note that because blockchain data cannot generally be changed, once a criminal steals a private key and publicly transfers the associated funds to another account, that transaction generally cannot be undone.

## 3.5 Ledgers

A *ledger* is a collection of transactions. Throughout history, pen and paper ledgers have been used to keep track of the exchange of goods and services. In modern times, ledgers have been stored digitally, often in large databases owned and operated by a centralized trusted third party (i.e., the owner of the ledger) on behalf of a community of users. These ledgers with centralized ownership can be implemented in a centralized or distributed fashion (i.e., just one server or a coordinating cluster of servers).

There is growing interest in exploring having distributed ownership of the ledger. Blockchain technology enables such an approach using both distributed ownership as well as a distributed physical architecture. The distributed physical architecture of blockchain networks often involve a much larger set of computers than is typical for centrally managed distributed physical architecture. The growing interest in distributed ownership of ledgers is due to possible trust, security, and reliability concerns related to ledgers with centralized ownership:

- Centrally owned ledgers may be lost or destroyed; a user must trust that the owner is properly backing up the system.

  o A blockchain network is distributed by design, creating many backup copies all updating and syncing to the same ledger data between peers. A key benefit to blockchain technology is that every user can maintain their own copy of the ledger. Whenever new full nodes join the blockchain network, they reach out to discover other full nodes and request a full copy of the blockchain network's ledger, making loss or destruction of the ledger difficult.
  Note – certain blockchain implementations provide the capability to support concepts such as private transactions or private channels. Private transactions facilitate the delivery of information only to those nodes participating in a transaction and not the entire network.

- Centrally owned ledgers may be on a homogeneous network, where all software, hardware and network infrastructure may be the same. Because of this characteristic, the overall system resiliency may be reduced since an attack on one part of the network will work on everywhere.

  o A blockchain network is a heterogeneous network, where the software, hardware and network infrastructure are all different. Because of the many differences between nodes on the blockchain network, an attack on one node is not guaranteed to work on other nodes.

- Centrally owned ledgers may be located entirely in specific geographic locations (e.g., all in one country). If network outages were to occur in that location, the ledger and services which depend on it may not be available.

  o A blockchain network can be comprised of geographically diverse nodes which may be found around the world. Because of this, and the blockchain network working in a peer-to-peer fashion, it is resilient to the loss of any node, or even an entire region of nodes.

- The transactions on a centrally owned ledger are not made transparently and may not be valid; a user must trust that the owner is validating each received transaction.

  o A blockchain network must check that all transactions are valid; if a malicious node was transmitting invalid transactions, others would detect and ignore them, preventing the invalid transactions from propagating throughout the blockchain network.

- The transaction list on a centrally owned ledger may not be complete; a user must trust that the owner is including all valid transactions that have been received.

  o A blockchain network holds all accepted transactions within its distributed ledger. To build a new block, a reference must be made to a previous block – therefore building on top of it. If a publishing node did not include a reference to the latest block, other nodes would reject it.

- The transaction data on a centrally owned ledger may have been altered; a user must trust that the owner is not altering past transactions.

- o A blockchain network utilizes cryptographic mechanisms such as digital signatures and cryptographic hash functions to provide tamper evident and tamper resistant ledgers.

- The centrally owned system may be insecure; a user must trust that the associated computer systems and networks are receiving critical security patches and have implemented best practices for security. The system may be breached and have had personal information stolen because of insecurities.

  - o A blockchain network, due to the distributed nature, provides no centralized point of attack. Generally, information on a blockchain network is publicly viewable, and offers nothing to steal. To attack blockchain network users, an attacker would need to individually target them. Targeting the blockchain itself would be met with the resistance of the honest nodes present in the system. If an individual node was not patched, it would only affect that node – not the system overall.

## 3.6 Blocks

Blockchain network users submit candidate transactions to the blockchain network via software (desktop applications, smartphone applications, digital wallets, web services, etc.). The software sends these transactions to a node or nodes within the blockchain network. The chosen nodes may be non-publishing full nodes as well as publishing nodes. The submitted transactions are then propagated to the other nodes in the network, but this by itself does not place the transaction in the blockchain. For many blockchain implementations, once a pending transaction has been distributed to nodes, it must then wait in a queue until it is added to the blockchain by a publishing node.

Transactions are added to the blockchain when a publishing node publishes a block. A *block* contains a block header and block data. The block header contains metadata for this block. The block data contains a list of validated and authentic transactions which have been submitted to the blockchain network. Validity and authenticity is ensured by checking that the transaction is correctly formatted and that the providers of digital assets in each transaction (listed in the transaction's 'input' values) have each cryptographically signed the transaction. This verifies that the providers of digital assets for a transaction had access to the private key which could sign over the available digital assets. The other full nodes will check the validity and authenticity of all transactions in a published block and will not accept a block if it contains invalid transactions.

It should be noted that every blockchain implementation can define its own data fields; however, many blockchain implementations utilize data fields like the following:

- Block Header

  - o The block number, also known as block height in some blockchain networks.

  - o The previous block header's hash value.

  - o A hash representation of the block data (different methods can be used to accomplish this, such as a generating a Merkle tree (defined in Appendix B), and storing the root hash, or by utilizing a hash of all the combined block data).

  - o A timestamp.

- o The size of the block.

- o The nonce value. For blockchain networks which utilize mining, this is a number which is manipulated by the publishing node to solve the hash puzzle (see Section 4.1 for details). Other blockchain networks may or may not include it or use it for another purpose other than solving a hash puzzle.

- Block Data

  - o A list of transactions and ledger events included within the block.

  - o Other data may be present.

## 3.7 Chaining Blocks

Blocks are chained together through each block containing the hash digest of the previous block's header, thus forming the *blockchain*. If a previously published block were changed, it would have a different hash. This in turn would cause all subsequent blocks to also have different hashes since they include the hash of the previous block. This makes it possible to easily detect and reject altered blocks. Figure 3 shows a generic chain of blocks.

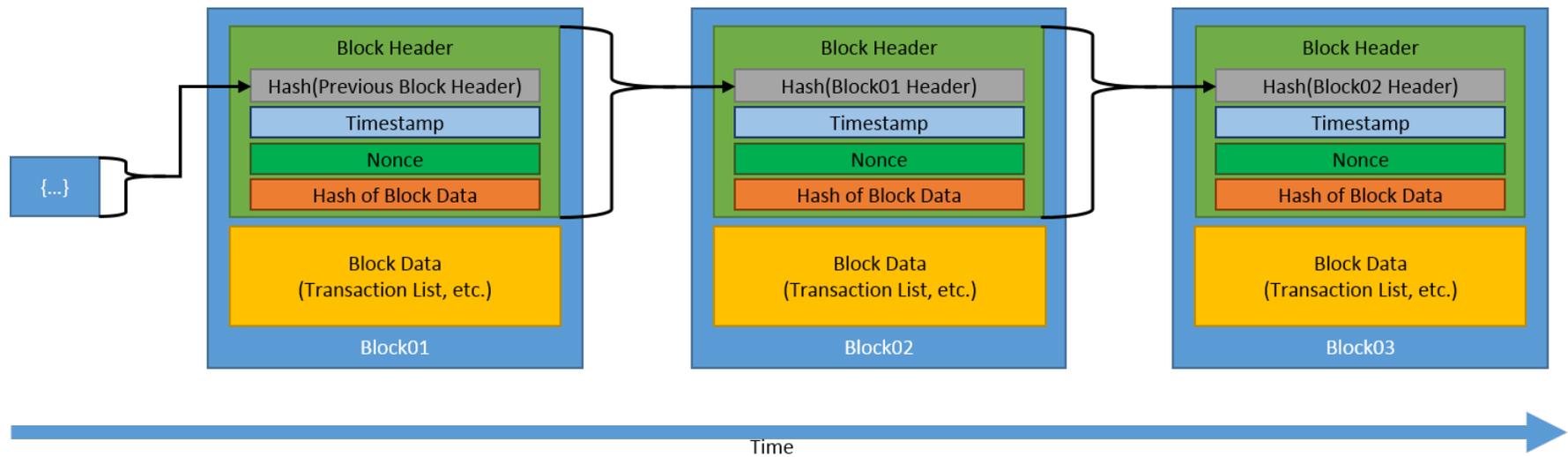

**Figure 3: Generic Chain of Blocks**

## 4    Consensus Models

A key aspect of blockchain technology is determining which user publishes the next block. This is solved through implementing one of many possible consensus models. For permissionless blockchain networks there are generally many publishing nodes competing at the same time to publish the next block. They usually do this to win cryptocurrency and/or transaction fees. They are generally mutually distrusting users that may only know each other by their public addresses. Each publishing node is likely motivated by a desire for financial gain, not the well-being of the other publishing nodes or even the network itself.

In such a situation, why would a user propagate a block that another user is attempting to publish? Also, who resolves conflicts when multiple nodes publish a block at approximately the same time? To make this work, blockchain technologies use *consensus models* to enable a group of mutually distrusting users to work together.

When a user joins a blockchain network, they agree to the initial state of the system. This is recorded in the only pre-configured block, the *genesis block*. Every blockchain network has a published genesis block and every block must be added to the blockchain after it, based on the agreed-upon consensus model. Regardless of the model, however, each block must be valid and thus can be validated independently by each blockchain network user. By combining the initial state and the ability to verify every block since then, users can independently agree on the current state of the blockchain. Note that if there were ever two valid chains presented to a full node, the default mechanism in most blockchain networks is that the 'longer' chain is viewed as the correct one and will be adopted; this is because it has had the most amount of work put into it. This happens frequently with some consensus models and will be discussed in detail.

The following properties are then in place:

- The initial state of the system is agreed upon (e.g., the genesis block).
- Users agree to the consensus model by which blocks are added to the system.
- Every block is linked to the previous block by including the previous block header's hash digest (except for the first 'genesis' block, which has no previous block and for which the hash of the previous block header is usually set to all zeros).
- Users can verify every block independently.

In practice, software handles everything and the users do not need to be aware of these details.

A key feature of blockchain technology is that there is no need to have a trusted third party provide the state of the system—every user within the system can verify the system's integrity. To add a new block to the blockchain, all nodes must come to a common agreement over time; however, some temporary disagreement is permitted. For permissionless blockchain networks, the consensus model must work even in the presence of possibly malicious users since these users might attempt to disrupt or take over the blockchain. Note that for permissioned blockchain networks legal remedies may be used if a user acts maliciously.

In some blockchain networks, such as permissioned, there may exist some level of trust between publishing nodes. In this case, there may not be the need for a resource intensive (computation time, investment, etc.) consensus model to determine which participant adds the next block to the chain. Generally, as the level of trust increases, the need for resource usage as a measure of generating trust decreases. For some permissioned blockchain implementations, the view of consensus extends beyond ensuring validity and authenticity of the blocks but encompasses the entire systems of checks and validations from the proposal of a transaction, to its final inclusion on a block.

In the following sections, several consensus models as well as the most common conflict resolution approach are discussed.

## 4.1    Proof of Work Consensus Model

In the proof of work (PoW) model, a user publishes the next block by being the first to solve a computationally intensive puzzle. The solution to this puzzle is the "proof" they have performed work. The puzzle is designed such that solving the puzzle is difficult but checking that a solution is valid is easy. This enables all other full nodes to easily validate any proposed next blocks, and any proposed block that did not satisfy the puzzle would be rejected.

A common puzzle method is to require that the hash digest of a block header be less than a target value. Publishing nodes make many small changes to their block header (e.g., changing the nonce) trying to find a hash digest that meets the requirement. For each attempt, the publishing node must compute the hash for the entire block header. Hashing the block header many times becomes a computationally intensive process. The target value may be modified over time to adjust the difficulty (up or down) to influence how often blocks are being published.

For example, Bitcoin, which uses the proof of work model, adjusts the puzzle difficulty every 2016 blocks to influence the block publication rate to be around once every ten minutes. The adjustment is made to the difficulty level of the puzzle, and essentially either increases or decreases the number of leading zeros required. By increasing the number of leading zeros, it increases the difficulty of the puzzle, because any solution must be less than the difficulty level – meaning there are fewer possible solutions. By decreasing the number of leading zeros, it decreases the difficulty level, because there are more possible solutions. This adjustment is to maintain the computational difficulty of the puzzle, and therefore maintain the core security mechanism of the Bitcoin network. Available computing power increases over time, as does the number of publishing nodes, so the puzzle difficulty is generally increasing.

Adjustments to the difficulty target aim to ensure that no entity can take over block production, but as a result the puzzle solving computations require significant resource consumption. Due to the significant resource consumption of some proof of work blockchain networks, there is a move to add publishing nodes to areas where there is a surplus supply of cheap electricity.

An important aspect of this model is that the work put into a puzzle does not influence one's likelihood of solving the current or future puzzles because the puzzles are independent. This means that when a user receives a completed and valid block from another user, they are incentivized to discard their current work and to start building off the newly received block instead because they know the other publishing nodes will be building off it.

As an example, consider a puzzle where, using the SHA-256 algorithm, a computer must find a hash value meeting the following target criteria (known as the difficulty level):

        SHA256("blockchain" + Nonce) = Hash Digest starting with "000000"

In this example, the text string "`blockchain`" is appended with a nonce value and then the hash digest is calculated. The nonce values used will be numeric values only. This is a relatively easy puzzle to solve and some sample output follows:

        SHA256("blockchain0") =
        0xbd4824d8ee63fc82392a6441444166d22ed84eaa6dab11d4923075975acab938
        (not solved)

        SHA256("blockchain1") =
        0xdb0b9c1cb5e9c680dfff7482f1a8efad0e786f41b6b89a758fb26d9e223e0a10
        (not solved)

        …

        SHA256("blockchain10730895") =
        0x000000ca1415e0bec568f6f605fcc83d18cac7a4e6c219a957c10c6879d67587
        (solved)

To solve this puzzle, it took 10,730,896 guesses (completed in 54 seconds on relatively old hardware, starting at 0 and testing one value at a time).

In this example, each additional "leading zero" value increases the difficulty. By increasing the target by one additional leading zero ("0000000"), the same hardware took 934,224,175 guesses to solve the puzzle (completed in 1 hour, 18 minutes, 12 seconds):

        SHA256("blockchain934224174") =
        0x0000000e2ae7e4240df80692b7e586ea7a977eacbd031819d0e603257edb3a81

There is currently no known shortcut to this process; publishing nodes must expend computation effort, time, and resources to find the correct nonce value for the target. Often the publishing nodes attempt to solve this computationally difficult puzzle to claim a reward of some sort (usually in the form of a cryptocurrency offered by the blockchain network). The prospect of being rewarded for extending and maintaining the blockchain is referred to as a reward system or incentive model.

Once a publishing node has performed this work, they send their block with a valid nonce to full nodes in the blockchain network. The recipient full nodes verify that the new block fulfills the puzzle requirement, then add the block to their copy of the blockchain and resend the block to their peer nodes. In this manner, the new block gets quickly distributed throughout the network of participating nodes. Verification of the nonce is easy since only a single hash needs to be done to check to see if it solves the puzzle.

For many proof of work based blockchain networks, publishing nodes tend to organize themselves into "pools" or "collectives" whereby they work together to solve puzzles and split the reward. This is possible because work can be distributed between two or more nodes across a collective to share the workload and rewards. Splitting the example program into quarters, each node can take an equal amount of the nonce value range to test:

- `Node 1: check nonce 0000000000 to 0536870911`
- `Node 2: check nonce 0536870912 to 1073741823`
- `Node 3: check nonce 1073741824 to 1610612735`
- `Node 4: check nonce 1610612736 to 2147483647`

The following result was the first to be found to solve the puzzle:

```
SHA256("blockchain1700876653") =
0x000000003ba55d20c9cbd1b6fb34dd81c3553360ed918d07acf16dc9e75d7c7f1
```

This is a completely new nonce, but still one that solved the puzzle. It took 90,263,918 guesses (completed in 10 minutes, 14 seconds). Dividing up the work amongst many more machines yields much better results, as well as more consistent rewards in a proof of work model.

The use of a computationally difficult puzzle helps to combat the "Sybil Attack" – a computer security attack (not limited to blockchain networks) where an attacker can create many nodes (i.e., creating multiple identities) to gain influence and exert control. The proof of work model combats this by having the focus of network influence being the amount of computational power (hardware, which costs money) mixed with a lottery system (the most hardware increases likelihood but does not guarantee it) versus in network identities (which are generally costless to create).

## 4.2    Proof of Stake Consensus Model

The proof of stake (PoS) model is based on the idea that the more stake a user has invested into the system, the more likely they will want the system to succeed, and the less likely they will want to subvert it. Stake is often an amount of cryptocurrency that the blockchain network user has invested into the system (through various means, such as by locking it via a special transaction type, or by sending it to a specific address, or holding it within special wallet software). Once staked, the cryptocurrency is generally no longer able to be spent. Proof of stake blockchain networks use the amount of stake a user has as a determining factor for publishing new blocks. Thus, the likelihood of a blockchain network user publishing a new block is tied to the ratio of their stake to the overall blockchain network amount of staked cryptocurrency.

With this consensus model, there is no need to perform resource intensive computations (involving time, electricity, and processing power) as found in proof of work. Since this consensus model utilizes fewer resources, some blockchain networks have decided to forego a block creation reward; these systems are designed so that all the cryptocurrency is already distributed among users rather than new cryptocurrency being generated at a constant pace. In such systems, the reward for block publication is then usually the earning of user provided transaction fees.

The methods for how the blockchain network uses the stake can vary. Here we discuss four approaches: random selection of staked users, multi-round voting, coin aging systems and delegate systems. Regardless of the exact approach, users with more stake are more likely to publish new blocks.

When the choice of block publisher is a random choice (sometimes referred to as *chain-based proof of stake*), the blockchain network will look at all users with stake and choose amongst them based on their ratio of stake to the overall amount of cryptocurrency staked. So, if a user had 42 % of the entire blockchain network stake they would be chosen 42 % of the time; those with 1 % would be chosen 1 % of the time.

When the choice of block publisher is a multi-round voting system (sometime referred to as *Byzantine fault tolerance proof of stake* [12]) there is added complexity. The blockchain network will select several staked users to create proposed blocks. Then all staked users will cast a vote for a proposed block. Several rounds of voting may occur before a new block is decided upon. This method allows all staked users to have a voice in the block selection process for every new block.

When the choice of block publisher is through a coin age system referred to as a *coin age proof of stake,* staked cryptocurrency has an *age* property. After a certain amount of time (such as 30 days) the staked cryptocurrency can *count* towards the owning user being selected to publish the next block. The staked cryptocurrency then has its *age* reset, and it cannot be used again until after the requisite time has passed. This method allows for users with more stake to publish more blocks, but to not dominate the system – since they have a cooldown timer attached to every cryptocurrency coin *counted* towards creating blocks. Older coins and larger groups of coins will increase the probability of being chosen to publish the next block. To prevent stakeholders from hoarding aged cryptocurrencies, there is generally a built-in maximum to the probability of winning.

When the choice of block publisher is through a delegate system, users vote for nodes to become publishing nodes – therefore creating blocks on their behalf. Blockchain network users' voting power is tied to their stake so the larger the stake, the more weight the vote has. Nodes who receive the most votes become publishing nodes and can validate and publish blocks. Blockchain network users can also vote against an established publishing node, to try to remove them from the set of publishing nodes. Voting for publishing nodes is continuous and remaining a publishing node can be quite competitive. The threat of losing publishing node status, and therefore rewards and reputation is constant so publishing nodes are incentivized to not act maliciously. Additionally, blockchain network users vote for delegates, who participate in the governance of the blockchain. Delegates will propose changes, and improvements, which will be voted on by blockchain network users.

It is worth noting that a problem known as "nothing at stake" may arise from some proof of stake algorithms. If multiple competing blockchains were to exist at some point (because of a temporary ledger conflict as discussed in Section 4.7), a staked user could act on every such competing chain – since it is essentially free to do so. The staked user may do this as a way of increasing their odds of earning a reward. This can cause multiple blockchain branches to continue to grow without being reconciled into a singular branch for extended periods of time.

Under proof of stake systems, the "rich" can more easily stake more of the digital assets, earning themselves more digital assets; however, to obtain the majority of digital assets within a system to "control" it is generally cost prohibitive.

## 4.3    Round Robin Consensus Model

Round Robin is a consensus model that is used by some permissioned blockchain networks. Within this model of consensus, nodes take turns in creating blocks. Round Robin Consensus has a long history grounded in distributed system architecture. To handle situations where a publishing node is not available to publish a block on its turn, these systems may include a time limit to enable available nodes to publish blocks so that unavailable nodes will not cause a halt in block publication. This model ensures no one node creates the majority of the blocks. It benefits from a straightforward approach, lacks cryptographic puzzles, and has low power requirements.

Since there is a need for trust amongst nodes, round robin does not work well in the permissionless blockchain networks used by most cryptocurrencies. This is because malicious nodes could continuously add additional nodes to increase their odds of publishing new blocks. In the worst case, they could use this to subvert the correct operation of the blockchain network.

## 4.4    Proof of Authority/Proof of Identity Consensus Model

The proof of authority (also referred to as proof of identity) consensus model relies on the partial trust of publishing nodes through their known link to real world identities. Publishing nodes must have their identities proven and verifiable within the blockchain network (e.g., identifying documents which have been verified and notarized and included on the blockchain). The idea is that the publishing node is staking its identity/reputation to publish new blocks. Blockchain network users directly affect a publishing node's reputation based on the publishing node's behavior. Publishing nodes can lose reputation by acting in a way that the blockchain network users disagree with, just as they can gain reputation by acting in a manner that the blockchain network users agree with. The lower the reputation, the less likelihood of being able to publish a block. Therefore, it is in the interest of a publishing node to maintain a high reputation. This algorithm only applies to permissioned blockchain networks with high levels of trust.

## 4.5    Proof of Elapsed Time Consensus Model

Within the proof of elapsed time (PoET) consensus model, each publishing node requests a wait time from a secure hardware time source within their computer system. The secure hardware time source will generate a random wait time and return it to the publishing node software. Publishing nodes take the random time they are given and become idle for that duration. Once a publishing node wakes up from the idle state, it creates and publishes a block to the blockchain network, alerting the other nodes of the new block; any publishing node that is still idle will stop waiting, and the entire process starts over.

This model requires ensuring that a random time was used, since if the time to wait was not selected at random a malicious publishing node would just wait the minimum amount of time by default to dominate the system. This model also requires ensuring that the publishing node waited the actual time and did not start early. These requirements are being solved by executing software in a trusted execution environment found on some computer processors (such as Intel's Software Guard Extensions[5], or AMD's Platform Security Processor[6], or ARM's TrustZone[7]).

Verified and trusted software can run in these secure execution environments and cannot be altered by outside programs. A publishing node would query software running in this secure environment for a random time and then wait for that time to pass. After waiting the assigned time, the publishing node could request a signed certificate that the publishing node waited the randomly assigned time. The publishing node then publishes the certificate along with the block.

---

[5] Intel SGX - https://software.intel.com/en-us/sgx

[6] AMD Secure Technology - https://www.amd.com/en/technologies/security

[7] ARM TrustZone - https://www.arm.com/products/silicon-ip-security

## 4.6    Consensus Comparison Matrix

| Name | Goals | Advantages | Disadvantages | Domains | Implementations |
|------|-------|------------|---------------|---------|-----------------|
| **Proof of work (PoW)** | To provide a barrier to publishing blocks in the form of a computationally difficult puzzle to solve to enable transactions between untrusted participants. | Difficult to perform denial of service by flooding network with bad blocks.<br><br>Open to anyone with hardware to solve the puzzle. | Computationally intensive (by design), power consumption, hardware arms race.<br><br>Potential for 51 % attack by obtaining enough computational power. | Permissionless cryptocurrencies | Bitcoin, Ethereum, many more |
| **Proof of stake (PoS)** | To enable a less computationally intensive barrier to publishing blocks, but still enable transactions between untrusted participants. | Less computationally intensive than PoW.<br><br>Open to anyone who wishes to stake cryptocurrencies.<br><br>Stakeholders control the system. | Stakeholders control the system.<br><br>Nothing to prevent formation of a pool of stakeholders to create a centralized power.<br><br>Potential for 51 % attack by obtaining enough financial power. | Permissionless cryptocurrencies | Ethereum Casper, Krypton |
| **Delegated PoS** | To enable a more efficient consensus model through a 'liquid democracy' where participants vote (using cryptographically signed messages) to elect and revoke the rights of delegates to validate and secure the blockchain. | Elected delegates are economically incentivized to remain honest<br><br>More computationally efficient than PoW | Less node diversity than PoW or pure PoS consensus implementations<br><br>Greater security risk for node compromise due to constrained set of operating nodes<br><br>As all delegates are 'known' there may an incentive for block producers to collude and accept bribes, compromising the security of the system | Permissionless cryptocurrencies<br><br>Permissioned Systems | Bitshares, Steem, Cardano, EOS |

| Name | Goals | Advantages | Disadvantages | Domains | Implementations |
|---|---|---|---|---|---|
| **Round Robin** | Provide a system for publishing blocks amongst approved/trusted publishing nodes | Low computational power.<br><br>Straightforward to understand. | Requires large amount of trust amongst publishing nodes. | Permissioned Systems | MultiChain |
| **Proof of Authority/Identity** | To create a centralized consensus process to minimize block creation and confirmation rate | Fast confirmation time<br><br>Allows for dynamic block production rates<br><br>Can be used in sidechains to blockchain networks which utilize another consensus model | Relies on the assumption that the current validating node has not been compromised<br><br>Leads to centralized points of failure<br><br>The reputation of a given node is subject to potential for high tail-risk as it could be compromised at any time. | Permissioned Systems, Hybrid (sidechain) Systems | Ethereum Kovan testnet, POA Chain, various permissioned systems using Parity |
| **Proof of Elapsed Time (PoET)** | To enable a more economic consensus model for blockchain networks, at the expense of deeper security guarantees associated with PoW. | Less computationally expensive than PoW | Hardware requirement to obtain time.<br><br>Assumes the hardware clock used to derive time is not compromised<br><br>Given speed-of-late latency limits, true time synchronicity is essentially impossible in distributed systems [13] | Permissioned Networks | Hyperledger Sawtooth |

## 4.7    Ledger Conflicts and Resolutions

As discussed previously, for some blockchain networks it is possible that multiple blocks will be published at approximately the same time. This can cause differing versions of a blockchain to exist at any given moment; these must be resolved quickly to have consistency in the blockchain network. In this section, we discuss how these situations are generally handled.

With any distributed network, some systems within the network will be behind on information or have alternative information. This depends on network latency between nodes and the proximity of groups of nodes. Permissionless blockchain networks are more prone to have conflicts due to their openness and number of competing publishing nodes. A major part of agreeing on the state of the blockchain network (coming to consensus) is resolving conflicting data.

For example:

- `node_A` creates `block_n(A)` with transactions #1, 2 and 3. `node_A` distributes it to some nodes.
- `node_B` creates `block_n(B)` with transactions #1, 2 and 4. `node_B` distributes it to some nodes.
- **There is a conflict.**
    - `block_n` will not be the same across the network.
        - `block_n(A)` contains transaction #3, but not transaction #4.
        - `block_n(B)` contains transaction #4, but not transaction #3.

Conflicts temporarily generate different versions of the blockchain, which is depicted in Figure 4. These differing versions are not "wrong"; rather, they were created with the information each node had available. The competing blocks will likely contain different transactions, so those with `block_n(A)` may see transfers of digital assets not present in `block_n(B)`. If the blockchain network deals with cryptocurrency, then a situation may occur where some cryptocurrency may both be spent and unspent, depending on which version of the blockchain is being viewed.

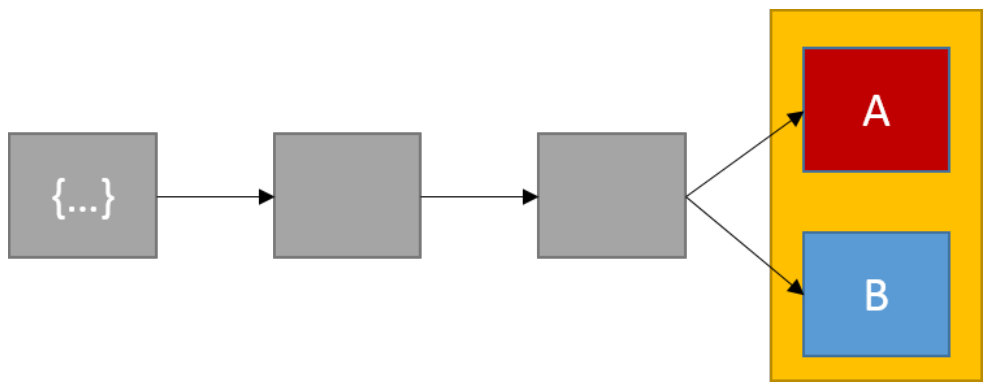

**Figure 4: Ledger in Conflict**

Conflicts are usually quickly resolved. Most blockchain networks will wait until the next block is published and use that chain as the "official" blockchain, thus adopting the "longer blockchain". As in Figure 5, the blockchain containing `block_n(B)` becomes the "official" chain, as it got the next valid block. Any transaction that was present in `block_n(A)`, the orphaned block, but not present in the `block_n(B)` chain, is returned to the pending transaction pool (which is where all transactions which have not been included within a block reside). Note that this set of pending transactions is maintained locally at each node as there is no central server in the architecture.

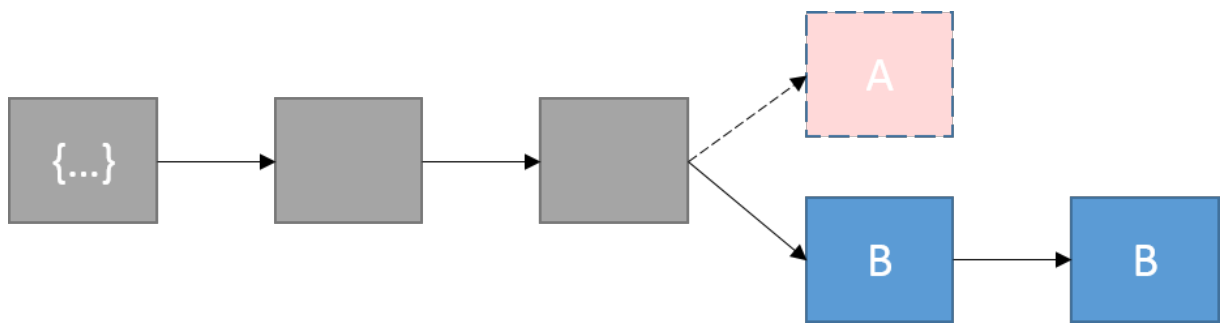

**Figure 5: The chain with block_n(B) adds the next block, the chain with block_n(A) is now orphaned**

Due to the possibility of blocks being overwritten, a transaction is not usually accepted as confirmed until several additional blocks have been created on top of the block containing the relevant transaction. The acceptance of a block is often probabilistic rather than deterministic since blocks can be superseded. The more blocks that have been built on top of a published block, the more likely it is that the initial block will not be overwritten.

Hypothetically, a node in a proof of work blockchain network with enormous amounts of computing power could start at the genesis block and create a longer chain than the currently existing chain, thereby wiping out the entire blockchain history. This does not happen in practice due to the prohibitively large amount of resources that this would require. Also, some blockchain implementations lock specific older blocks within the blockchain software by creating checkpoints to ensure that this can never happen.

## 5      Forking

Performing changes and updating technology can be difficult at the best of times. For permissionless blockchain networks which are comprised of many users, distributed around the world, and governed by the consensus of the users, it becomes extremely difficult. Changes to a blockchain network's protocol and data structures are called *forks*. They can be divided into two categories: *soft forks* and *hard forks*. For a soft fork, these changes are backwards compatible with nodes that have not been updated. For a hard fork, these changes are not backwards compatible because the nodes that have not been updated will reject the blocks following the changes. This can lead to a split in the blockchain network creating multiple versions of the same blockchain. Permissioned blockchain networks, due to the publishing nodes and users being known, can mitigate the issues of forking by requiring software updates.

Note that the term fork is also used by some blockchain networks to describe temporary ledger conflicts (e.g., two or more blocks within the blockchain network with the same block number) as described in Section 4.7. While this is a fork in the ledger, it is temporary and does not stem from a software change.

### 5.1    Soft Forks

A *soft fork* is a change to a blockchain implementation that is backwards compatible. Non-updated nodes can continue to transact with updated nodes. If no (or very few) nodes upgrade, then the updated rules will not be followed.

An example of a soft fork occurred on Bitcoin when a new rule was added to support escrow[8] and time-locked refunds. In 2014, a proposal was made to repurpose an operation code that performed no operation (OP_NOP2) to CHECKLOCKTIMEVERIFY, which allows a transaction output to be made spendable at a point in the future [14]. For nodes that implement this change, the node software will perform this new operation, but for nodes that do not support the change, the transaction is still valid, and execution will continue as if a NOP [9] had been executed.

A fictional example of a soft fork would be if a blockchain decided to reduce the size of blocks (for example from 1.0 MB to 0.5 MB). Updated nodes would adjust the block size and continue to transact as normal; non-updated nodes would see these blocks as valid – since the change made does not violate their rules (i.e., the block size is under their maximum allowed). However, if a non-updated node were to create a block with a size greater than 0.5 MB, updated nodes would reject them as invalid.

### 5.2    Hard Forks

A *hard fork* is a change to a blockchain implementation that is not backwards compatible. At a given point in time (usually at a specific block number), all publishing nodes will need to switch to using the updated protocol. Additionally, all nodes will need to upgrade to the new protocol so that they do not reject the newly formatted blocks. Non-updated nodes cannot continue to transact on the updated blockchain because they are programmed to reject any block that does not follow their version of the block specification.

---

[8] Funds placed into a third party to be disseminated based on conditions (via multi-signature transactions)

[9] NOP meaning No Operation

Publishing nodes that do not update will continue to publish blocks using the old format. User nodes that have not updated will reject the newly formatted blocks and only accept blocks with the old format. This results in two versions of the blockchain existing simultaneously. Note that users on different hard fork versions cannot interact with one another. It is important to note that while most hard forks are intentional, software errors may produce unintentional hard forks.

A well-known example of a hard fork is from Ethereum. In 2016, a smart contract was constructed on Ethereum called the Decentralized Autonomous Organization (DAO). Due to flaws in how the smart contract was constructed, an attacker extracted Ether, the cryptocurrency used by Ethereum, resulting in the theft of $50 million [15]. A hard fork proposal was voted on by Ether holders, and the clear majority of users agreed to hard fork and create a new version of the blockchain, without the flaw, and that also returned the stolen funds.

With cryptocurrencies, if there is a hard fork and the blockchain splits then users will have independent currency on both forks (having double the number of coins in total). If all the activity moves to the new chain, the old one may eventually not be used since the two chains are not compatible (they will be independent currency systems). In the case of the Ethereum hard fork, the clear majority of support moved to the new fork, the old fork was renamed Ethereum Classic and continued operating.

## 5.3   Cryptographic Changes and Forks

If flaws are found in the cryptographic technologies within a blockchain network, the only solution may be to create a hard fork, depending on the significance of the flaw. For example, if a flaw was found in the underlying algorithms, there could be a fork requiring all future clients to use a stronger algorithm. Switching to a new hashing algorithm could pose a significant practical problem because it could invalidate all existing specialized mining hardware.

Hypothetically, if SHA-256 were discovered to have a flaw, blockchain networks that utilize SHA-256 would need a hard fork to migrate to a new hash algorithm. The block that switched over to the new hash algorithm would "lock" all previous blocks into SHA-256 (for verification), and all new blocks would need to utilize the new hashing algorithm. There are many cryptographic hash algorithms, and blockchain networks can make use of whichever suits their needs. For example, while Bitcoin uses SHA-256, Ethereum uses Keccak-256 [8].

One possibility for the need to change cryptographic features present in a blockchain network would be the development of a practical quantum computer system, which would be capable of greatly weakening (and in some cases, rendering useless) existing cryptographic algorithms. NIST Internal Report (NISTIR) 8105, Report on Post-Quantum Cryptography [16] provides a table describing the impact of quantum computing on common cryptographic algorithms. Table 2 replicates this table.

**Table 2: Impact of Quantum Computing on Common Cryptographic Algorithms**

| Cryptographic Algorithm | Type | Purpose | Impact from Large-Scale Quantum Computer |
|---|---|---|---|
| AES | Symmetric key | Encryption | Larger key sizes needed |
| SHA-2, SHA-3 | N/A | Hash functions | Larger output needed |
| RSA | Public key | Signatures, key establishment | No longer secure |
| ECDSA, ECDH (Elliptic Curve Cryptography) | Public key | Signatures, key exchange | No longer secure |
| DSA (Finite Field Cryptography) | Public key | Signatures, key exchange | No longer secure |

The cryptographic algorithms utilized within most blockchain technologies for asymmetric-key pairs will need to be replaced if a powerful quantum computer becomes a reality. This is because algorithms that rely on the computational complexity of integer factorization (such as RSA) or work on solving discrete logarithms (such as DSA and Diffie-Hellman) are very susceptible to being broken by quantum computing. The hashing algorithms used by blockchain networks are much less susceptible to quantum computing attacks but are still weakened.

## 6     Smart Contracts

The term smart contract dates to 1994, defined by Nick Szabo as "a computerized transaction protocol that executes the terms of a contract. The general objectives of smart contract design are to satisfy common contractual conditions (such as payment terms, liens, confidentiality, and even enforcement), minimize exceptions both malicious and accidental, and minimize the need for trusted intermediaries." [17].

Smart contracts extend and leverage blockchain technology. A *smart contract* is a collection of code and data (sometimes referred to as functions and state) that is deployed using cryptographically signed transactions on the blockchain network (e.g., Ethereum's smart contracts, Hyperledger Fabric's chaincode). The smart contract is executed by nodes within the blockchain network; all nodes that execute the smart contract must derive the same results from the execution, and the results of execution are recorded on the blockchain.

Blockchain network users can create transactions which send data to public functions offered by a smart contract. The smart contract executes the appropriate method with the user provided data to perform a service. The code, being on the blockchain, is also tamper evident and tamper resistant and therefore can be used (among other purposes) as a trusted third party. A smart contract can perform calculations, store information, expose properties to reflect a publicly exposed state and, if appropriate, automatically send funds to other accounts. It does not necessarily even have to perform a financial function. For example, the authors of this document have created an Ethereum smart contract that publicly generate trustworthy random numbers [18]. It is important to note that not every blockchain can run smart contracts.

The smart contract code can represent a multi-party transaction, typically in the context of a business process. In a multi-party scenario, the benefit is that this can provide attestable data and transparency that can foster trust, provide insight that can enable better business decisions, reduce costs from reconciliation that exists in traditional business to business applications, and reduce the time to complete a transaction.

Smart contracts must be deterministic, in that given an input they will always produce the same output based on that input. Additionally, all the nodes executing the smart contract must agree on the new state that is obtained after the execution. To achieve this, smart contracts cannot operate on data outside of what is directly passed into it (e.g., smart contracts cannot obtain web services data from within the smart contract – it would need to be passed in as a parameter). Any smart contract which uses data from outside the context of its own system is said to use an 'Oracle' (the oracle problem is described in section 7.3).

For many blockchain implementations, the publishing nodes execute the smart contract code simultaneously when publishing new blocks. There are some blockchain implementations in which there are publishing nodes which do not execute smart contract code, but instead validate the results of the nodes that do.  For smart contract enabled permissionless blockchain networks (such as Ethereum) the user issuing a transaction to a smart contract will have to pay for the cost of the code execution. There is a limit on how much execution time can be consumed by a call to a smart contract, based on the complexity of the code. If this limit is exceeded, execution stops, and the transaction is discarded. This mechanism not only rewards the publishers for executing the smart contract code, but also prevents malicious users from deploying and then accessing smart contracts that will perform a denial of service on the publishing nodes by consuming all resources (e.g., using infinite loops).

For smart contract enabled permissioned blockchain networks, such as those utilizing Hyperledger Fabric's chaincode, there may not be a requirement for users to pay for smart contract code execution. These networks are designed around having known participants, and other methods of preventing bad behavior can be employed (e.g., revoking access).

## 7    Blockchain Limitations and Misconceptions

There is a tendency to overhype and overuse most nascent technology. Many projects will attempt to incorporate the technology, even if it is unnecessary. This stems from the technology being relatively new and not well understood, the technology being surrounded by misconceptions, and the fear of missing out. Blockchain technology has not been immune. This section highlights some of the limitations and misconceptions of blockchain technology.

### 7.1   Immutability

Most publications on blockchain technology describe blockchain ledgers as being immutable. However, this is not strictly true. They are tamper evident and tamper resistant which is a reason they are trusted for financial transactions. They cannot be considered completely immutable, because there are situations in which the blockchain can be modified. In this section we will look at different ways in which the concept of immutability for blockchain ledgers can be violated.

The chain of blocks itself cannot be considered completely immutable. For some blockchain implementations, the most recently published, or 'tail' blocks are subject to being replaced (by a longer, alternative chain with different 'tail' blocks). As noted earlier, most blockchain networks use the strategy of adopting the longest chain (the one with the most amount of work put into it) as truth when there are multiple competing chains. If two chains are competing, but each include their own unique sequence of tail blocks, whichever is longer will be adopted. However, this does not mean that the transactions within the replaced blocks are lost – rather they may have been included in a different block or returned to the pending transaction pool. This degree of weak immutability for tail blocks is why most blockchain network users wait several block creations before considering a transaction to be valid.

For permissionless blockchain networks, the adoption of a longer, alternate chain of blocks could be the result of a form of attack known as a 51 % attack [19]. For this, the attacker simply garners enough resources to outpace the block creation rate of rest of the blockchain network (holding more than 51 % of the resources applied towards producing new blocks). Depending on the size of the blockchain network, this could be a very cost prohibitive attack carried out by state level actors [20]. The cost to perform this type of attack increases the further back in the blockchain the attacker wishes to make a change. This attack is not technically difficult (e.g., it is just repeating the normal process of the blockchain implementation, but with selected transactions either included or omitted, and at a faster pace), it is just expensive.

For permissioned blockchain networks, this attack can be mitigated. There is generally an owner or consortium of blockchain network users who allow publishing nodes to join the blockchain network and remove publishing nodes from the blockchain network, which gives them a great amount of control.  There is less likely to be competing chains since the owner or consortium can force publishing nodes to collaborate fairly since non-cooperating publishing nodes can simply have their privileges removed. There are likely additional legal contracts in place for the blockchain network users which may include clauses for misconduct and the ability to take legal action. While this control is useful to prevent misconduct, it means that any number of blocks can be replaced through legitimate methods if desired by the owner or consortium.

## 7.2    Users Involved in Blockchain Governance

The governance of blockchain networks deals with the rules, practices and processes by which the blockchain network is directed and controlled. A common misconception is that blockchain networks are systems without control and ownership. The phrase "no one controls a blockchain!" is often exclaimed. This is not strictly true. Permissioned blockchain networks are generally setup and run by an owner or consortium, which governs the blockchain network. Permissionless blockchain networks are often governed by blockchain network users, publishing nodes, and software developers. Each group has a level of control that affects the direction of the blockchain network's advancement.

Software developers create the blockchain software that is utilized by a blockchain network. Since most blockchain technologies are open source, it is possible to inspect the source code, and compile it independently; it is even possible to create separate but compatible software as a means of bypassing pre-compiled software released by developers. However, not every user will have the ability to do this, which means that the developer of the blockchain software will play a large role in the blockchain network's governance. These developers may act in the interest of the community at large and are held accountable. For example, in 2013 Bitcoin developers released a new version of the most popular Bitcoin client which introduced a flaw and started two competing chains of blocks. The developers had to decide to either keep the new version (which had not yet been adopted by everyone) or revert to the old version [21]. Either choice would result in one chain being discarded—and some blockchain network user's transactions becoming invalid. The developers made a choice, reverted to the old version, and successfully controlled the progress of the Bitcoin blockchain.

This example was an unintentional fork; however, developers can purposely design updates to blockchain software to change the blockchain protocol or format. With enough user adoption, a successful fork can be created. Such forks of blockchain software updates are often discussed at length and coordinated with the involved users. For permissionless blockchain networks, this is usually the publishing nodes. There is often a long discussion and adoption period before an event occurs where all users must switch to the newly updated blockchain software at some chosen block to continue recording transactions on the new "main" fork.

For permissionless blockchain networks, although the developers maintain a large degree of influence, users can reject a change by the developers by refusing to install updated software. Of the blockchain network users, the publishing nodes have significant control since they create and publish new blocks. The user base usually adopts the blocks produced by the publishing nodes but is not required to do so. An interesting side effect of this is that permissionless blockchain networks are essentially ruled by the publishing nodes and may marginalize a segment of users by forcing them to adopt changes they may disagree with to stay with the main fork.

For permissioned blockchain networks, control and governance is driven by members of the associated owner or consortium. The consortium can govern who can join the network, when members are removed from the network, coding guidelines for smart contracts, etc.

In summary, the software developers, publishing nodes, and blockchain network users all play a part in the blockchain network governance.

## 7.3    Beyond the Digital

Blockchain networks work extremely well with the data within their own digital systems. However, when they need to interact with the real world, there are some issues (often called the Oracle Problem [22]). A blockchain network can be a place to record both human input data as well as sensor input data from the real world, but there may be no method to determine if the input data reflects real world events. A sensor could be malfunctioning and recording data that is inaccurate. Humans could record false information (intentionally or unintentionally). These issues are not specific to blockchain networks, but to digital systems overall. However, for blockchain networks that are pseudonymous, dealing with data misrepresentation outside of the digital network can be especially problematic.

For example, if a cryptocurrency transaction took place to purchase a real-world item there is no way to determine within the blockchain network whether the shipment took place, without relying on outside sensor or human input.

Many projects have attempted to address the 'Oracle problem' and create reliable mechanisms to ingest external data in a way that is both trustworthy and accurate.  For example, projects like 'Oraclize' provide mechanisms to take web API data and convert it into blockchain readable byte/opcode. Within the context of decentralized applications, these projects may be considered centralized as they provide single points of failure for attackers to compromise. As a result, projects like 'Mineable Oracle Contract' [23] have recently arisen to enable oracle ingestion in a way that is inspired by blockchain technology and built atop established consensus models and economic incentives.

## 7.4    Blockchain Death

Traditional centralized systems are created and taken down constantly, and blockchain networks will likely not be different. However, because they are decentralized, there is a chance that when a blockchain network "shuts down" it will never be fully shut down, and that there may always be some lingering blockchain nodes running.

A defunct blockchain would not be suitable for a historical record, since without many publishing nodes, a malicious user could easily overpower the few publishing nodes left and redo and replace any number of blocks.

## 7.5    Cybersecurity

The use of blockchain technology does not remove inherent cybersecurity risks that require thoughtful and proactive risk management. Many of these inherent risks involve a human element. Therefore, a robust cybersecurity program remains vital to protecting the network and participating organizations from cyber threats, particularly as hackers develop more knowledge about blockchain networks and their vulnerabilities.

Existing cybersecurity standards and guidance remain highly relevant for ensuring the security of systems that interface and/or rely on blockchain networks. Subject to certain adjustments to consider specific attributes of blockchain technology, existing standards and guidance provide a strong foundation for protecting blockchain networks from cyberattacks.

In addition to general principles and controls, there are specific cybersecurity standards with relevance to blockchain technology which already exist and are in wide use by many industries. For instance, the NIST Cybersecurity Framework expressly states that it is "not a one-size-fits-all approach to managing cybersecurity risk" because "organizations will continue to have unique risks—different threats, different vulnerabilities, different risk tolerances—and how they implement the practices in the [Framework] will vary." With that said, even though the Framework was not designed for blockchain technology specifically, its standards are broad enough to cover blockchain technology and to help institutions develop policies and processes that identify and control risks affecting blockchain technology.

### 7.5.1 Cyber and Network-based Attacks

Blockchain technologies are touted as being extremely secure due to the tamper evident and tamper resistant design – once a transaction is committed to the blockchain, it generally cannot be changed. However, this is only true for transactions which have been included in a published block. Transactions that have not yet been included in a published block within the blockchain are vulnerable to several types of attacks. For blockchain networks which have transactional timestamps, spoofing time or adjusting the clock of a member of an ordering service could have positive or negative effects on a transaction, making time and the communication of time an attack vector. Denial of service attacks can be conducted on the blockchain platform or on the smart contract implemented on the platform.

Blockchain networks and their applications are not immune to malicious actors who can conduct network scanning and reconnaissance to discover and exploit vulnerabilities and launch zero-day attacks. In the rush to deploy blockchain-based services, newly coded applications (like smart contracts) may contain new and known vulnerabilities and deployment weaknesses that will be discovered and then attacked through the network just like how websites or applications are attacked today.

## 7.6 Malicious Users

While a blockchain network can enforce transaction rules and specifications, it cannot enforce a user code of conduct. This is problematic in permissionless blockchain networks, since users are pseudonymous and there is not a one-to-one mapping between blockchain network user identifiers and users of the system. Permissionless blockchain networks often provide a reward (e.g., a cryptocurrency) to motivate users to act fairly; however, some may choose to act maliciously if that provides greater rewards. The largest problem for malicious users is getting enough power (be it a stake in the system, processing power, etc.) to cause damage. Once a large enough malicious collusion is created, malicious mining actions can include:

- Ignoring transactions from specific users, nodes, or even entire countries.
- Creating an altered, alternative chain in secret, then submitting it once the alternative chain is longer than the real chain. The honest nodes will switch to the chain that has the most "work" done (per the blockchain protocol). This could attack the principle of a blockchain network being tamper evident and tamper resistant [24].
- Refusing to transmit blocks to other nodes, essentially disrupting the distribution of information (this is not an issue if the blockchain network is sufficiently decentralized).

While malicious users can be annoyances and create short-term harm, blockchain networks can perform hard forks to combat them. Whether damages done (money lost) would be reversed would be up to the developers and users of the blockchain network.

In addition to there being malicious users of the network, the administrators of the infrastructure for permissioned blockchain networks may also act maliciously. For example, an infrastructure administrator may be able (depending upon the exact configuration) to take over block production, exclude certain users from performing transactions, rewrite block history, double spend coin, delete resources, or re-route or block network connections.

## 7.7    No Trust

Another common misinterpretation comes from people hearing that there is no "trusted third party" in a blockchain and assuming blockchain networks are "trustless" environments. While there is no trusted third party certifying transactions in permissionless blockchain networks (in permissioned systems it is less clear, as administrators of those systems act as an administrator of trust by granting users admission and permissions), there is still a great deal of trust needed to work within a blockchain network:

- There is trust in the cryptographic technologies utilized. For example, cryptographic algorithms or implementations can have flaws.
- There is trust in the correct and bug free operation of smart contracts, which might have unintended loopholes and flaws.
- There is trust in the developers of the software to produce software that is as bug-free as possible.
- There is trust that most users of the blockchain are not colluding in secret. If a single group or individual can control more than 50 percent of all block creation power, it is possible to subvert a permissionless blockchain network. However, generally obtaining the necessary computational power is prohibitively expensive.
- For blockchain network users not running a full node, there is trust that nodes are accepting and processing transactions fairly.

## 7.8    Resource Usage

Blockchain technology has enabled a worldwide network where every transaction is verified and the blockchain is kept in sync amongst a multitude of users. For blockchain networks utilizing proof of work, there are many publishing nodes expending large amounts of processing time and, more importantly, consuming a lot of electricity. A proof of work method is an effective solution for "hard to solve, easy to verify" proofs; however, it generally requires significant resource usage.  Because of their different applications, and trust models, many permissioned blockchain technologies do not use a resource intensive proof, but rather they utilize different mechanisms to achieve consensus.

The proof of work consensus model is designed for the case where there is little to no trust amongst users of the system. It ensures that publishing nodes cannot game the system[10] by always being able to solve the puzzles and thereby control the blockchain and the transactions added to it. However, a major concern surrounding the proof of work consensus model is its use of energy in solving the puzzles.

---

[10]    Use the rules and procedures meant to protect the system to manipulate the system for a desired result.

The amount of energy used is often not trivial; for example, some estimate that currently the Bitcoin blockchain network uses around the same amount of electricity as the entire country of Ireland [25]. It has also been speculated that the Bitcoin blockchain network will consume as much electricity as the entire country of Denmark by 2020 [26][27][28]. Software and hardware will continue to improve, resulting in more efficient puzzle solving (reducing the amount of electricity utilized) [29]. However, blockchain networks are also still growing, resulting in harder puzzle difficulty.

An additional strain on resources occurs whenever a new full node is created; the node must obtain (usually through downloading) most of or all the blockchain data (Bitcoin's blockchain data is over 175 gigabytes and growing as of this writing) [30]. This process uses a lot of network bandwidth.

## 7.9 Inadequate Block Publishing Rewards

A potential limitation is the risk of inadequate rewards for publishing a block. The combination of increased competition, increased computational resources needed to have meaningful contributions to pools of publishing nodes, and highly volatile market prices in the cryptocurrency market creates the risk that the expected return for any given cryptocurrency may be less than the power costs needed to run publishing node software. Thus, the expected return for other cryptocurrencies may be more attractive.

Cryptocurrencies that are not able to consistently and adequately reward publishing nodes risk delays in publishing blocks and processing transactions. These delays could therefore reduce confidence in the cryptocurrency, reducing its market value further. It could then become increasingly less attractive for publishing nodes to contribute to that cryptocurrency's publishing efforts. Even worse, such weakened cryptocurrencies open themselves up to being attacked by nodes with large amounts of resources that may maliciously alter the blockchain or deny service to users attempting to submit transactions.

## 7.10 Public Key Infrastructure and Identity

When hearing that blockchain technology incorporates a public key infrastructure, some people immediately believe it intrinsically supports identity. This is not the case, as there may not be a one-to-one relationship of private key pairs to users (a user can have multiple private keys), nor is there a one-to-one relationship between blockchain addresses and public keys (multiple addresses can be derived from a single public key).

Digital signatures are often used to prove identity in the cybersecurity world, and this can lead to confusion about the potential application of a blockchain to identity management. A blockchain's transaction signature verification process links transactions to the owners of private keys but provides no facility for associating real-world identities with these owners. In some cases, it is possible to connect real-world identities with private keys, but these connections are made through processes outside, and not explicitly supported by, the blockchain. For example, a law enforcement agency could request records from an exchange that would connect transactions to specific individuals. Another example is an individual posting a cryptocurrency address on their personal website or social media page for donations, this would provide a link from address to real world identity.

While it is possible to use blockchain technology in identity management frameworks that require a distributed ledger component, it is important to understand that typical blockchain implementations are not designed to serve as standalone identity management systems. There is more to having secure digital identities than simply implementing a blockchain.

## 8    Application Considerations

Since blockchain technology is still new, a lot of organizations are looking at ways to incorporate it into their businesses. The fear of missing out on this technology is quite high, and most organizations approach the problem as "we want to use blockchain somewhere, where can we do that?" which leads to frustrations with the technology as it cannot be applied universally. A better approach would be to first understand blockchain technology, where it fits, and then identify systems (new and old) that may fit the blockchain paradigm.

Blockchain technology solutions may be suitable if the activities or systems require features such as:

- Many participants
- Distributed participants
- Want or need for lack of trusted third party
- Workflow is transactional in nature (e.g., transfer of digital assets/information between parties)
- A need for a globally scarce digital identifier (i.e., digital art, digital land, digital property)
- A need for a decentralized naming service or ordered registry
- A need for a cryptographically secure system of ownership
- A need to reduce or eliminate manual efforts of reconciliation and dispute resolutions
- A need to enable real time monitoring of activity between regulators and regulated entities
- A need for full provenance of digital assets and a full transactional history to be shared amongst participants

Several agencies and organizations have developed guides to help determine if a blockchain is suitable for a particular system or activity, and which kind of blockchain technology would be of most benefit. In this section, some articles and advice are highlighted from several different sectors – federal government, academia, technical publications, technology websites, and software developers.

The United States Department of Homeland Security (DHS) Science & Technology Directorate has been investigating blockchain technology and has created a flowchart to help one determine whether a blockchain may be needed for a development initiative. The flowchart is reproduced here, with permission.

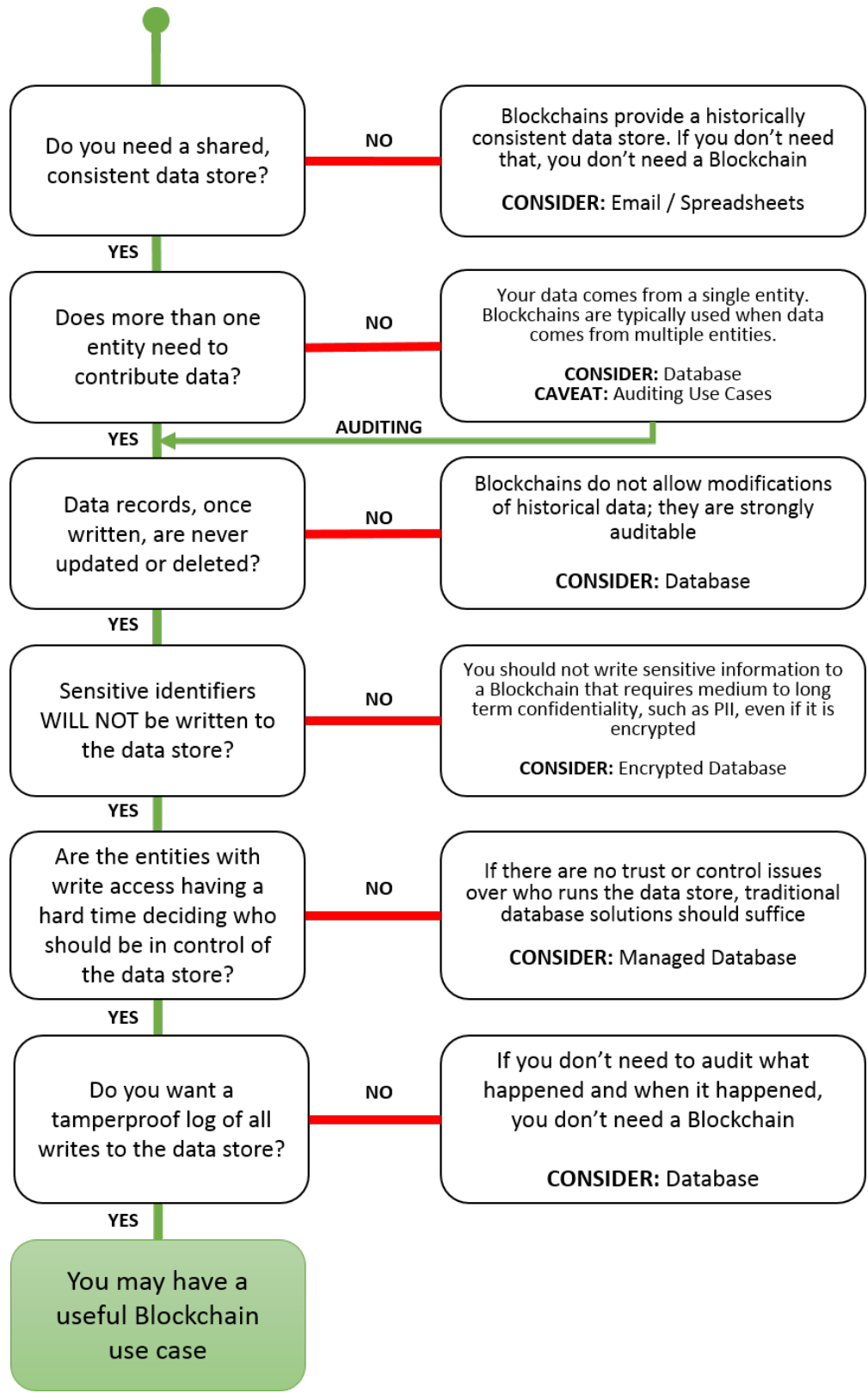

**Figure 6 - DHS Science & Technology Directorate Flowchart**

The American Council for Technology and Industry Advisory Council (ACT-IAC) have been developing both a blockchain technology primer and a blockchain playbook. ACT-IAC is a public/private partnership which facilitates collaboration and discussion between government and industry experts. ACT-IAC has developed a blockchain primer document [31], which aims to provide an overview of the technology. A second document, a blockchain playbook [32], provides a set of questions with weights to help organizations in their consideration of the technology.

There is no lack of whitepapers and news articles with a title like "Do you need a blockchain?" Two computer scientists at the Eidgenössische Technische Hochschule (ETH) Zürich university in Switzerland wrote a whitepaper titled "*Do you need a Blockchain?*" [33] which provides the background, properties, and a critical view on several use cases. Although not created by the authors, a website [34] has implemented the flowchart presented in the paper in an interactive form. However, examining the flowchart logic, as well as website code, most paths lead to "no" with only a few leading to "maybe." This critical view on the technology is one that most organizations should take; organizations should examine whether existing technologies can better solve their problems.

The Institute of Electrical and Electronics Engineers (IEEE) published in their *Spectrum* magazine the article "*Do you need a blockchain?*"[35]. The article emphasizes the utility a blockchain may provide (as an anti-censorship tool), but also discusses the tradeoff that must be made by moving away from a traditional system. Removal of trusted third parties means relying on multiple sources of "unaffiliated participants" acting in coordination, which depending on the type of blockchain platform, may be difficult to govern. The article also discusses that the technology is changing at a rapid pace – so it is difficult to predict where it will end up in a few years' time. The article includes a flowchart of its own to help the reader decide whether they need a blockchain. Finally, the article ends with the following statement: *"But you should also consider the possibility that you don't need a blockchain at all."* This is pertinent to those who may be desperately looking to include blockchain in their organization's portfolio.

Technology sites are also asking organizations to look closely at the technology and apply it only when necessary. Coindesk, a technology website specializing in cryptocurrency and blockchain news, technical matters and editorials, has written the article "*Don't use a blockchain unless you really need one*"[36]. The article gives some small examples about how most data today is owned by siloed organizations, and that as users we only supply it to them. It asks what the world would look like if users owned all their data. The article makes the point that the largest benefit of blockchain technology is its decentralization and can be summed up with the article's most critical point: "Despite some of the hype, blockchains are 'incredibly inefficient,' Ravikant said. 'It's worth paying the cost when you need the decentralization, but it's not when you don't.'"

Even software developers are urging organizations to examine the key aspects of the technology and how it could be applied to a problem. One such developer wrote on the website C# Corner the article "*Do You Need A Blockchain*" [37]. This article touches on the history of blockchain technology and brings to light a primary reason for the use of blockchain technology: "Blockchain brings trust to a transactional system."

By utilizing a blockchain cryptographic trust can be introduced into a previously no to low trust system. The article goes on to ask several pointed questions (and provides a flowchart) for helping to decide whether a blockchain network would be of benefit.

While several sources have been mentioned above for deciding if a blockchain would be applicable, there are many more. Most of the advice surrounding blockchain technology is: investigate it and use it if it is appropriate – not because it is new.

## 8.1    Additional Blockchain Considerations

When deciding whether to utilize a blockchain, one must take into consideration additional factors and determine if these factors limit one's ability to use a blockchain or a particular type of blockchain:

- **Data Visibility**
  - o  Permissioned blockchain networks may or may not reveal blockchain data publicly. The data may only be available to those within the blockchain network. Consider scenarios where data may be governed by policy or regulations (such as Personally Identifiable Information (PII) or General Data Protection Regulation (GDPR) regulations). Data such as this may or may not be appropriate to store even within a permissioned blockchain network.
  - o  Permissionless blockchain networks can allow anyone to inspect and contribute to the blockchain. The data is generally public. This leads to several questions that must be considered. Does the data for the application need to be available to everyone? Is there any harm to having public data?
- **Full transactional history** – Some blockchain networks provide a full public history of a digital asset – from creation, to every transaction it is included in. This feature may be beneficial for some solutions, and not beneficial for others.
- **Fake Data Input** – Since multiple users are contributing to a blockchain, some could submit false data, mimicking data from valid sources (such as sensor data). It is difficult to automate the verification of data that enters a blockchain network. Smart contract implementations may provide additional checks to help validate data where possible.
- **Tamper evident and tamper resistant data** – Many applications follow the "CRUD" (create, read, update, delete) functions for data. With a blockchain, there is only "CR" (create, read). There are methods that can be employed to "deprecate" older data if a newer version is found, but there is no removal process for the original data. By using new transactions to amend and update previous transactions, data can be updated while providing a full history. However, even if a new transaction marked an older transaction as "deleted" – the data would still be present in the blockchain data, even if it is not shown within an application processing the data.
- **Transactions Per Second** – Transaction processing speed is highly dependent on the consensus model used. Currently transactions on many permissionless blockchain networks are not executed at the same pace as other information technology solutions due to a slow publication time for blocks (usually in terms of seconds, but sometimes minutes). Thus, some slowdown in blockchain dependent applications may occur while waiting for data to be posted. One must ask if their application can handle relatively slow transaction processing?

- **Compliance** – The use of blockchain technology does not exclude a system from following any applicable laws and regulations. For example, there are many compliance considerations with regards to legislation and policies tied to PII or GDPR that identify that certain information should not be placed on the blockchain. In addition, certain countries may limit the type of data that can be transferred across its geographic boundary. In other instances, certain legislation may dictate that the "first write" of financial transactions must be written to a node which is present within their borders. In any of these cases, a public, permissionless chain may be less appropriate, with a permissioned or hybrid approach required to satisfy regulatory needs.

  An additional example of laws and regulations are for any blockchain network which manages federal records. Federal records are subject to many laws and regulations.[11] Federal agencies themselves must follow specific federal guidelines when utilizing blockchain technology.[12]

- **Permissions** – For permissioned blockchain networks, there are considerations around the permissions themselves
    - Granularity – do the permissions within the system allow for enough granularity for specific roles that users may need (in a manner like Role-Based Access Control methods) to perform actions within the system
        - Permissioned blockchain networks allow for more traditional roles such as administrator, user, validator, auditor, etc.
    - Administration – who can administer permissions? Once permissions are administered to a user, can they easily be revoked?

- **Node Diversity** – A blockchain network is only as strong as the aggregate of all the existing nodes participating in the network. If all the nodes share similar hardware, software, geographic location, and messaging schema then there exists a certain amount of risk associated with the possibility of undiscovered security vulnerabilities. This risk is mitigated through the decentralization of the network of heterogeneous devices, which may be defined as "the non-shared characteristics between any one node and the generalized set"

---

[11] Such as found in the National Archives and Records Administration handbook https://www.archives.gov/records-mgmt/handbook/records-mgmt-language.html

[12] Such as found in the National Archives and Administration policy guide https://www.archives.gov/records-mgmt/policy/universalermrequirements

## 9    Conclusions

Blockchain technology is a new tool with potential applications for organizations, enabling secure transactions without the need for a central authority. Starting in 2009[13], with Bitcoin leveraging blockchain technology, there has been an increasing number of blockchain technology-based solutions.

The first applications were electronic cash systems with the distribution of a global ledger containing all transactions. These transactions are secured with cryptographic hashes, and transactions are signed and verified using asymmetric-key pairs. The transaction history efficiently and securely records a chain of events in a way that any attempt to edit or change a past transaction will also require a recalculation of all subsequent blocks of transactions.

The use of blockchain technology is still in its early stages, but it is built on widely understood and sound cryptographic principles. Currently, there is a lot of hype around the technology, and many proposed uses for it. Moving forward, it is likely that the hype will die down, and blockchain technology will become just another tool that can be used.

As detailed throughout this publication, a blockchain relies on existing network, cryptographic, and recordkeeping technologies but uses them in a new manner. It will be important that organizations are able to look at the technologies and both the advantages and disadvantages of using them. Once a blockchain is implemented and widely adopted, it may become difficult to change it. Once data is recorded in a blockchain, that data is usually there forever, even when there is a mistake. Applications that utilize the blockchain as a data layer work around the fact that the actual blockchain data cannot be altered by making later blocks and transactions act as updates or modifications to earlier blocks and transactions. This software abstraction allows for modifications to working data, while providing a full history of changes. For some organizations these are desirable features. For others, these may be deal breakers preventing the adoption of blockchain technology.

Blockchain technology is still new and organizations should treat blockchain technology like they would any other technological solution at their disposal--use it only in appropriate situations.

---

[13] Although the whitepaper *Bitcoin: A Peer-to-Peer Electronic Cash System* was published in 2008, the actual Bitcoin network would not launch until 2009.

## Appendix A—Acronyms

Selected acronyms and abbreviations used in this paper are defined below.

| | |
|---|---|
| ACM | Association for Computing Machinery |
| ACT-IAC | American Council for Technology and Industry Advisory Council |
| ASIC | Application-Specific Integrated Circuit |
| BCH | Bitcoin Cash |
| BFT | Byzantine Fault Tolerant |
| BTC | Bitcoin |
| CPU | Central Processing Unit |
| CR | Create, Read |
| CRUD | Create, Read, Update, Delete |
| DAG | Directed Acyclic Graph |
| DAO | Decentralized Autonomous Organization |
| DHS | Department of Homeland Security |
| DID | Decentralized Identifier |
| DSA | Digital Signature Algorithm |
| ECDSA | Elliptic Curve Digital Signature Algorithm |
| ETC | Ethereum Classic |
| ETH | Ethereum |
| EVM | Ethereum Virtual Machine |
| FIPS | Federal Information Processing Standard |
| FOIA | Freedom of Information Act |
| GDPR | General Data Protection Regulation |
| GPU | Graphics Processing Unit |
| I2P | Invisible Internet Project |
| IEEE | Institute of Electrical and Electronics Engineers |
| IoT | Internet of Things |
| IR | Internal Report |
| ITL | Information Technology Laboratory |
| KYC | Know Your Customer |
| LDAP | Lightweight Directory Access Protocol |
| NARA | National Archives and Records Administration |
| NIST | National Institute of Standards and Technology |
| NISTIR | National Institute of Standards and Technology Internal Report |
| MB | Megabyte |
| PII | Personally Identifiable Information |
| PoET | Proof of Elapsed Time |
| PoS | Proof of Stake |
| PoW | Proof of Work |
| QR | Quick Response |
| RIPEMD | RACE Integrity Primitives Evaluation Message Digest |
| RSA | Rivest-Shamir-Adleman |
| SegWit | Segregated Witness |
| SHA | Secure Hash Algorithm |
| XRP | Ripple |

## Appendix B—Glossary

Selected terms used in this paper are defined below.

| | |
|---|---|
| Address | A short, alphanumeric string derived from a user's public key using a hash function, with additional data to detect errors. Addresses are used to send and receive digital assets. |
| Assets | Anything that can be transferred. |
| Asymmetric-key cryptography | A cryptographic system where users have a private key that is kept secret and used to generate a public key (which is freely provided to others). Users can digitally sign data with their private key and the resulting signature can be verified by anyone using the corresponding public key.<br><br>Also known as Public-key cryptography. |
| Block | A data structure containing a block header and block data. |
| Block data | The portion of a block that contains a set of validated transactions and ledger events. |
| Block header | The portion of a block that contains information about the block itself (block metadata), typically including a timestamp, a hash representation of the block data, the hash of the previous block's header, and a cryptographic nonce (if needed). |
| Block reward | A reward (typically cryptocurrency) awarded to publishing nodes for successfully adding a block to the blockchain. |
| Blockchain | Blockchains are distributed digital ledgers of cryptographically signed transactions that are grouped into blocks. Each block is cryptographically linked to the previous one (making it tamper evident) after validation and undergoing a consensus decision. As new blocks are added, older blocks become more difficult to modify (creating tamper resistance). New blocks are replicated across copies of the ledger within the network, and any conflicts are resolved automatically using established rules. |

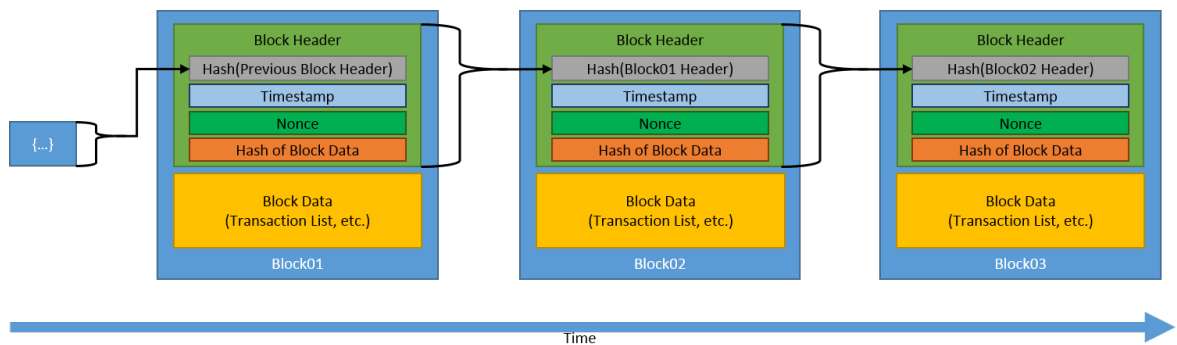

| | |
|---|---|
| Blockchain network user | Any single person, group, business, or organization which is using or operating a blockchain node. |

| Byzantine fault tolerant proof of stake consensus model | A proof of stake consensus model where the blockchain decides the next block by allowing all staked members to "vote" on which submitted block to include next. |
|---|---|
| Centralized network | A network configuration where participants must communicate with a central authority to communicate with one another. Since all participants must go through a single centralized source, the loss of that source would prevent all participants from communicating. 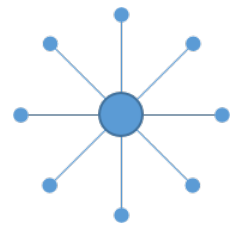 |
| Chain-based proof of stake consensus model | A proof of stake consensus model where the blockchain network decides the next block through pseudo-random selection, based on a personal stake to overall system asset ratio. |
| Checksum | Value computed on data to detect error or manipulation. |
| Confirmed | State of a transaction or block when consensus has been reached about its status of inclusion into the blockchain. |
| Conflict | One or more participants disagree on the state of the system. |
| Conflict resolution | A predefined method for coming to a consensus on the state of the system. For example, when portions of the system participants claim there is `State_A` and the rest of the participants claim there is `State_B`, there is a conflict. The system will automatically resolve this conflict by choosing the "valid" state as being the one from whichever group adds the next block of data. Any transactions "lost" by the state not chosen are added back into the pending transaction pool. |

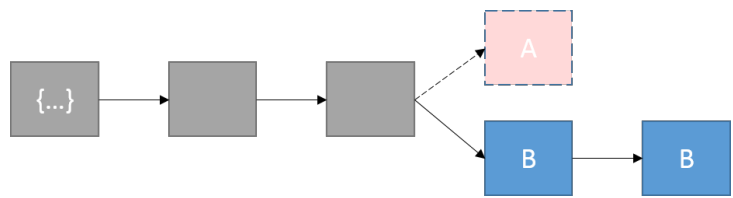

| Consensus model | A process to achieve agreement within a distributed system on the valid state. |
|---|---|
|  | Also known as a *consensus algorithm, consensus mechanism, consensus method.* |
| Cryptocurrency | A digital asset/credit/unit within the system, which is cryptographically sent from one blockchain network user to another. In the case of cryptocurrency creation (such as the reward for mining), the publishing node includes a transaction sending the newly created cryptocurrency to one or more blockchain network users. |
|  | These assets are transferred from one user to another by using digital signatures with asymmetric-key pairs. |
| Cryptographic hash function | A function that maps a bit string of arbitrary length to a fixed-length bit string. Approved hash functions satisfy the following properties: |

1. (*Preimage resistant*) It is computationally infeasible to compute the correct input value given some output value (the hash function is 'one way').
2. (*Second preimage resistant*) One cannot find an input that hashes to a specific output.
3. (*Collision resistant*) It is computationally infeasible to find any two distinct inputs that map to the same output.

See the NIST SP 800-175B Guideline for Using Cryptographic Standards in the Federal Government: Cryptographic Mechanisms, http://dx.doi.org/10.6028/NIST.SP.800-175B.

| | |
|---|---|
| Cryptographic nonce | An arbitrary number that is used once. |
| Decentralized network | A network configuration where there are multiple authorities that serve as a centralized hub for a subsection of participants. Since some participants are behind a centralized hub, the loss of that hub will prevent those participants from communicating. |

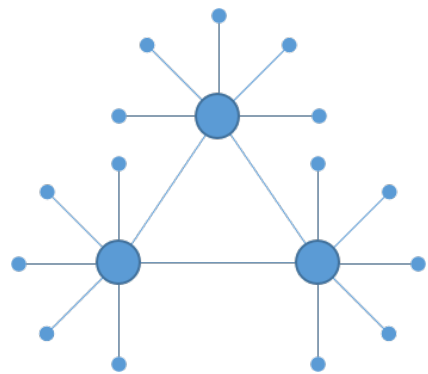

| | |
|---|---|
| Digest | See hash digest |
| Digital asset | Any asset that is purely digital, or is a digital representation of a physical asset |
| Digital signature | A cryptographic technique that utilizes asymmetric-keys to determine authenticity (i.e., users can verify that the message was signed with a private key corresponding to the specified public key), non-repudiation (a user cannot deny having sent a message) and integrity (that the message was not altered during transmission). |
| Distributed network | A network configuration where every participant can communicate with one another without going through a centralized point. Since there are multiple pathways for communication, the loss of any participant will not prevent communication.<br><br>This is also known as a peer-to-peer network. |

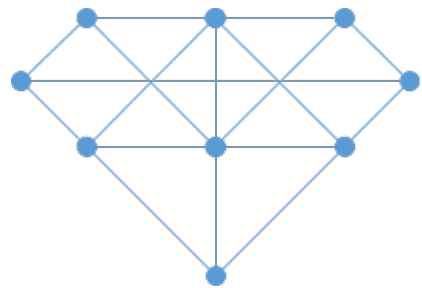

| | |
|---|---|
| Double spend (problem) | Transacting with the same set of digital assets more than once. This is a problem which has plagued many digital money systems, and a problem that most blockchain networks are designed to prevent. |
| Double spend (attack) | An attack where a blockchain network user attempts to explicitly double spend a digital asset. |
| Fault tolerance | A property of a system that allows proper operation even if components fail. |
| Fork | A change to blockchain network's software (usually the consensus algorithm). The changes may be backwards compatible - see Soft Fork, or the changes may not be backwards compatible - see Hard Fork. |
| Full node | A blockchain node that stores the blockchain data, passes along the data to other nodes, and ensures that newly added blocks are valid and authentic. |
| Genesis block | The first block of a blockchain network; it records the initial state of the system. |
| Hard fork | A change to a blockchain implementation that is not backwards compatible. Non-updated nodes cannot continue to transact with updated nodes. |
| Hash chain | An append-only data structure where data is bundled into data blocks that include a hash of the previous data block's data within the newest data block. This data structure provides evidence of tampering because any modification to a data block will change the hash digest recorded by the following data block. |
| Hash digest | The output of a hash function (e.g., hash(data) = digest). Also known as a message digest, digest or hash value. |
| Hash rate | The number of cryptographic hash functions a processor can calculate in a given time, usually denominated as hashes per second. |
| Hash value | See Hash digest. |
| Hashing | A method of calculating a relatively unique output (called a *hash digest*) for an input of nearly any size (a file, text, image, etc.) by applying a cryptographic hash function to the input data. |
| Immutable | Data that can only be written, not modified or deleted. |
| Incentive mechanism | See Reward system |
| Ledger | A record of transactions. |
| Lightweight node | A blockchain node that does not need to store a full copy of the blockchain and often passes its data to full nodes to be processed. |

| Merkle tree | A data structure where the data is hashed and combined until there is a singular root hash that represents the entire structure. |

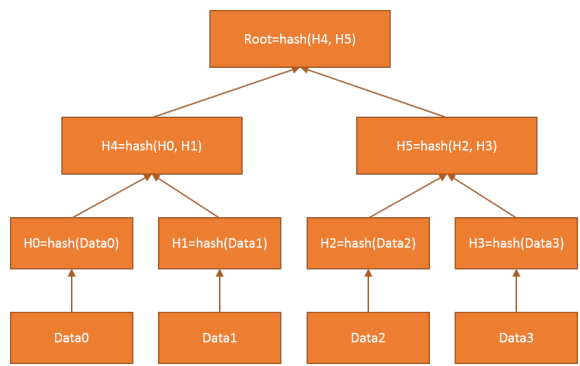

| Mining | The act of solving a puzzle within a proof of work consensus model. |
| Pending transaction pool | A distributed queue where candidate transactions wait until they are added to the blockchain. |
| | Also known as memory pool, or mempool. |
| Publishing node | A node that, in addition to all responsibilities required of a full node, is tasked with extending the blockchain by creating and publishing new blocks. |
| | Also known as a mining node, committing node, minting node. |
| Node | An individual system within the blockchain network. |
| Nonce | See Cryptographic Nonce |
| Orphan block | Any block that is not in the main chain after a temporary ledger conflict. |
| Permissioned | A system where every node, and every user must be granted permissions to utilize the system (generally assigned by an administrator or consortium). |
| Permissionless | A system where all users' permissions are equal and not set by any administrator or consortium. |
| Permissions | Allowable user actions (e.g., read, write, execute). |
| Proof of stake consensus model | A consensus model where the blockchain network is secured by users locking an amount of cryptocurrency into the blockchain network, a process called *staking*. Participants with more stake in the system are more likely to want it to succeed and to not be subverted, which gives them more weight during consensus. |
| Proof of work consensus model | A consensus model where a publishing node wins the right to publish the next block by expending time, energy, and computational cycles to solve a hard-to-solve, but easy-to-verify problem (e.g., finding the nonce which, when combined with the data to be added to the block, will result in a specific output pattern). |
| Public key cryptography | See Asymmetric-key cryptography. |

| Reward system | A means of providing blockchain network users an award for activities within the blockchain network (typically used as a system to reward successful publishing of blocks).<br><br>Also known as incentive system. |
| --- | --- |
| Round robin consensus model | A consensus model for permissioned blockchain networks where nodes are pseudo-randomly selected to create blocks, but a node must wait several block-creation cycles before being chosen again to add another new block. This model ensures that no one participant creates the majority of the blocks, and it benefits from a straightforward approach, lacking cryptographic puzzles, and having low power requirements. |
| Smart contract | A collection of code and data (sometimes referred to as functions and state) that is deployed using cryptographically signed transactions on the blockchain network. The smart contract is executed by nodes within the blockchain network; all nodes must derive the same results for the execution, and the results of execution are recorded on the blockchain. |
| Soft fork | A change to a blockchain implementation that is backwards compatible. Non-updated nodes can continue to transact with updated nodes. |
| Tamper evident | A process which makes alterations to the data easily detectable. |
| Tamper resistant | A process which makes alterations to the data difficult (hard to perform), costly (expensive to perform), or both. |
| Transaction | A recording of an event, such as the transfer of assets (digital currency, units of inventory, etc.) between parties, or the creation of new assets. |
| Transaction fee | An amount of cryptocurrency charged to process a blockchain transaction. Given to publishing nodes to include the transaction within a block. |
| Turing complete | A system (computer system, programming language, etc.) that can be used for any algorithm, regardless of complexity, to find a solution. |
| Wallet | Software used to store and manage asymmetric-keys and addresses used for transactions. |

## Appendix C—References